\documentclass[twocolumn,trackchanges]{aastex61}
\usepackage{amsmath,bm}
\usepackage{natbib}
\usepackage{xpatch}
\submitjournal{ApJ}

\begin{document}
\title{Test particle energization and the anisotropic effects of dynamical MHD turbulence}

\correspondingauthor{Carlos Gonz\'alez}
\email{caangonzalez@df.uba.ar}

\author{C.A. Gonz\'alez}
\affiliation{Departamento de F\'isica, Facultad de Ciencias Exactas y Naturales, Universidad de Buenos Aires and IFIBA,
CONICET, Ciudad universitaria, 1428 Buenos Aires, Argentina}

\author{P. Dmitruk}
\affiliation{Departamento de F\'isica, Facultad de Ciencias Exactas y
  Naturales, Universidad de Buenos Aires and IFIBA, CONICET, Ciudad
  universitaria, 1428 Buenos Aires, Argentina}
\author{P.D. Mininni}
\affiliation{Departamento de F\'isica, Facultad de Ciencias Exactas y
  Naturales, Universidad de Buenos Aires and IFIBA, CONICET, Ciudad
  universitaria, 1428 Buenos Aires, Argentina}
\author{W.H. Matthaeus}
\affiliation{Bartol Research Institute and Department of Physics and
  Astronomy, University of Delaware, Newark, Delaware, USA}



\begin{abstract}
In this paper we analyze the effect of dynamical three-dimensional MHD
turbulence on test particle acceleration, and compare how this
evolving system affects particle energization by current sheets
interaction, against frozen-in-time fields. To do this we analize the
ensamble particle acceleration for static electromagnetic fields
extracted from direct numerical simulations of the MHD equations, and
compare with the dynamical fields. We show that a reduction in
particle acceleration in the dynamical model results from the particle
trapping in the field lines, which forces the particles to remain in
a moving current sheet that suppress the longer exposure at the strong electric field gradients located
between structures, which is an efficient particle acceleration mechanism. In addition, we analize the
effect of anisotropy caused by the mean magnetic field. It is well known that
for sufficiently strong external fields, the system suffers a 
transition towards a two-dimensional flow. This causes an increment in
the size of the coherent structures, resulting in a magnetized state of the 
particles and the reduction of the particle energization.
\end{abstract}



\section{\label{sec:level1}INTRODUCTION:} 

The production of energetic particles in the Earth-Sun enviroment and
in the  interstellar medium brings the open question of what are the
mechanisms behind the generation of their non-thermal charged particle
population(\citet{Parker_1958}). In all these systems the flows are turbulent, as the
turbulent plasma state is observed in almost all astrophysical and
space physics systems. Acceleration of charged particles by turbulence
in the solar corona is also one of the candidates to explain coronal
heating, as well as the origin of solar wind energetic particles
observed at {\it in-situ} mesurements and at ground-based 
observatories(\citet{McCommas2007}).

As a result, many authors have studied and reported charged particle
acceleration by plasma turbulence (\citet{M2,L1}). In general, in these
problems turbulence covers a huge range of spatio-temporal scales,
from low frequency events well described by fluid plasma models (such
as magnetohydrodynamics or MHD, or Hall-MHD), up to very
high-frequency scales related to electron dynamics which are well
described by purely kinetic approaches.

In order to capture at least a fraction of the range of scales
involved in plasma turbulence, test particle simulations in
MHD flows have been used as a way to undestand how the macroscopic
behavior of a plasma could itself generate plasma heating and particle
energization. This problem has been treated in different ways, which
can be divided into two broad approaches. The first one consists on
modeling the turbulence as a random collection of
waves(\citet{Chandran2003,CH1,CHO1,Lynn2013}), a representation that
cannot retain the coherent structures formation and evolution, and
which in particular can  play a crucial role in particle energization
(\citet{M2,Tessein2015}). In the second approach, the turbulence is
generated as a result of the normal evolution of the MHD fluid
equations (\citet{PD1,hall,lehe2009heating,Bogdan2014,Dalena2014,Weidl2015}), and as such
it self-consistently includes the evolution of coherent
structures. However, test particle studies have been mostly carried
out using static fields obtained from the direct numerical simulation
of the MHD equations (i.e., the system is evolved in time to reach a
turbulent regime, and then a frozen-in-time snapshot of electric and magnetic fields is
used to compute particle acceleration). The problem with this method
is that it ignores the evolution of the coherent structures, and that
it does not allow for the presence of waves. 

Although there are some works that deal with the influence of dynamic MHD turbulence on test 
particles(\citet{lehe2009heating,Bogdan2014}), those papers does not make a complete analysis of
the dynamical fields effect which is in principle our goal in this paper. Besides, the comparision between static
and dynamic cases allow us to determine what is the most relevant mechanism for particle heating 
between particle-wave resonance or the interaction with well developed coherent structures in the 
turbulent flow.

Then, the first aim of this paper is therefore to look at the effect of
dynamical turbulence evolution on particle energization. To this end,
we performed different direct numerical simulations of MHD turbulence,
together with solving the equations for test particles with a
gyroradius of the order of the MHD dissipation scale, first in static
and then in dynamic electromagretic fields, and for different
values of a mean external magnetic field $B_0$. 

We found a reduction of particle energization in the dynamical case, 
and that this reduction is the result of particle trapping into the
dynamic current sheets. This trapping reduces the exposure time in the strong electric 
field gradient regions at the interface between current sheets, that is essentially needed for particle
energization. In spite of this reduction, the results validate
previously reported particle acceleration mechanisms in MHD turbulence
(\citet{M2,PD1,Bogdan2014,Dalena2014,Gonzalez2016_1}), and it
shows the importance of coherent structures in particle acceleration
phenomena.

The second aim of this paper is also to quantify the role of the
mean external field $B_0$ in coherent structure formation and particle
acceleration, as in many cases in space physics particle acceleration
takes place in the presence of a guide field. To do this, we performed
a set of simulations varying from an isotropic case ($B_0=0$) up
to a strong anisotropic case with a mean field $B_0=8$ (in units of
the fluctuating magnetic field ${\bf b}$). We show that the anisotropy
induced by the mean magnetic field produces large coherent structures
as the mean field increases. Consequently, for large magnetic field
the particles remain magnetized, that is, attached closely to magnetic field 
lines, 
and then cannot transport across the current sheets, decreasing
the posibility to stay long time in regions where the electric field gradient 
is strong, between current structures. The particle propagation along
the magnetic field lines then suppress the acceleration due to perpendicular 
electric field gradients between current sheet structures.

The organization of this paper is as follows: In section 2 we describe
the model used in our investigations, the equations and properties of
turbulent MHD fields, and the test particle model including the
parameters that relate particles and fields. In section 3.A we
present a comparision between the static and dynamical cases for an
isotropic case with $B_0=0$ and for an anisotropic case with
$B_0=2$. In section 3.B we discuss the effect of the mean magnetic
field on coherent structures and particle acceleration. Finally, in
section 4 we discuss our findings and present our conclusions.

\section{\label{sec:level2}MODELS:}

The macroscopic description of the plasma adopted here is modeled by
the three-dimensional compressible MHD equations: the continuity
(density) equation, the equation of motion, the magnetic field
induction equation, and the equation of state. These are given
respectively by Equations.~(1-4), which involve fluctuations  of the
velocity field $\textbf{u}$, magnetic field $\textbf{b}$, and density
$\rho$. We assume a large-scale background magnetic field $B_0$ in the
$z$-direction, so that the total magnetic field is 
$\mathbf{B = B_0 + b}$ with ${\bf B}_0 = B_0 \hat{z}$,
\begin{equation}
 \frac{\partial \rho}{\partial t} + \nabla \cdot (\textbf{u}\rho) = 0,
\end{equation}
\begin{equation}
 \frac{d\textbf{u}}{d t} = - \frac{\nabla p}{\rho} + \frac{\textbf{J} \times
   \textbf{B}}{4\pi\rho}
 + \nu \left( \nabla^2 \textbf{u} +    \frac{\nabla \nabla \cdot
     \textbf{u} }{3} \right) + {\bf F} ,
\end{equation}
\begin{equation}
\frac{\partial \textbf{B}}{\partial t} = \nabla \times (\textbf{u}
  \times \textbf{B}) + \eta \nabla^2 \textbf{B} + 
  \nabla \times \bf{\varepsilon} ,
\end{equation}
\begin{equation}
 \frac{p}{\rho^{\gamma}} = {\rm constant}.
\end{equation}
Here $p$ is the pressure, $\nu$ the viscosity, $\eta$ the magnetic
diffusivity, $\textbf{J}=\nabla \times \textbf{B} $ is the current
density, ${\bf F}$ is an external mechanical force, and 
$\bf{\varepsilon}$ is an external electromotive force. 
We assume a polytropic equation of state 
$p/p_0=(\rho/\rho_0)^{\gamma}$, with $\gamma=5/3$, where $p_0$ and
$\rho_0$ are respectively the equilibrium (reference) pressure and
density. The Hall current is 
not taken into account in equation.(3) but will be (nominally) later included 
on the particle motion equations through the generalized Ohm's law for
the electric field. The reason for that is that the dynamics of the 
$\textbf{u},\textbf{B}$ fields described by equations (1)-(4) is not importantly
affected by the presence of the Hall term, provided that the Hall scale (see below) is close to
the dissipation scale (see \citet{hall}).

The magnetic and velocity fields here are expressed in Alfv\'en speed
units based on field fluctuations. This Alfv\'en speed based on field
fluctuations is defined as $v_0=\sqrt{\left<b^2\right>/4\pi\rho_0}$. 
A characteristic plasma velocity can also be given by the parallel
Alfv\'en wave velocity along the mean magnetic field 
$v_A = B_0/\sqrt{4\pi\rho_0}$; this is proportional to the amplitude
of guide field. The ratio of fluid equilibrium pressure $p_0$ to
magnetic pressure $B_0^2$, the so-called $\beta$ of the plasma, is 
$\beta =p_0/B_0^2=1/(M B_0)^2$. The sonic Mach number $M=v_0/C_s$
relates the mean velocity field with the sound speed  
$C_s = \sqrt{\gamma p_0/\rho_0}$. We use the isotropic MHD turbulence
correlation length $L$ as a characteristic length (also called the
energy containing scale), defined as 
$L = L_{box} \int (E(k)/k) dk / \int E(k) dk$, where $E(k)$ is the energy
spectral density at wavenumber $k$, and $L_{box}$ is the linear size
of the domain. The unit timescale $t_0$, also called the eddy turnover
time, is derived from the unit length and from the fluctuation Alfven
speed $t_0=L/v_0$.

The MHD equations are solved numerically using a Fourier
pseudospectral method with periodic boundary conditions in a
normalized cube of size  $L_{box}=2\pi$; this scheme ensures exact
energy conservation for the continuous-time spatially-discrete
equations in the ideal ($\nu = \eta =0$) and not forced 
(${\bf F} = \bm{\varepsilon} = 0$) case (\citet{ghost}). The discrete
time integration is done with a high-order Runge-Kutta method, and a
resolution of $256^3$ Fourier modes is used. For the kinematic
Reynolds number $R=v_0L/\nu$ and the magnetic Reynolds number
$R_m=v_0L/\eta$, we take $R=R_m= 1000$, which are limited here by the
available spatial resolution. The Mach number in our simulations is
$M=0.25$, so we consider a weak compressible case.

In \citet{Gonzalez2016_1}. We reported the effect of
the compressibility of the flow on particle acceleration. In that
study we considered decaying turbulence from an initial perturbation, 
and after the turbulence was fully developed, the test particles were
injected into the system and evolved in a frozen-in-time snapshot of
the turbulent electromagnetic field. As in this paper we are
interested in studying the effect of dynamically evolving turbulence,
to mantain energy fluctuating around a mean value (i.e., to reach a
turbulent steady state) we must force the system externally using the 
mechanical and electromotive forces ${\bf F}$ and $\bm{\varepsilon}$
in Equations.~(2) and (3).

To this end we started the system from initially null magnetic and
velocity fields and forced until to get a quasi-stationary MHD state. The forcing
scheme that we employed (for both ${\bf F}$ 
and $\bm{\varepsilon}$) in all the
simulations showed here is based on that presented in \citet{pouquet1978}. The 
forcing introduces (on the 
average) zero mechanical and magnetic helicity, and zero
cross-correlation between the velocity and magnetic field
fluctuations. We used slowly-evolving random-phases forcings in 
the Fourier $k$-shells with $3\leq k \leq4$, with a correlation time 
$\tau = t_0$. In other words, in each eddy turnover time new random
(and uncorrelated) forcing functions ${\bf F}$ and $\bm{\varepsilon}$
were generated, and each forcing was linearly interpolated in time
from the previous random state to the new one in a time $t_0$. In this
way we prevented introducing sudden changes in the field that may
arise when delta-correlated in time forcing is used, and which may
affect the evolution of test particles.

Each random snapshot of the forcing functions was generated as
follows. For a forcing ${\bf f}$ (which may be either ${\bf F}$ or
$\bm{\varepsilon}$), we generated first two random
fields in Fourier
space,
\begin{equation}
{\bf v}_{j}^{(1)}({\bf k})=A(k)e^{i \phi_j}, \hspace{.5cm} 
{\bf v}_{j}^{(2)}({\bf k})=A(k)e^{i \psi_j},
\end{equation} 
where $j=1,2,3$ are the field Cartesian components, $\phi_j({\bf k})$
and $\psi_j ({\bf k})$ are random phases, and $A(k)$ is 1 if $3\leq k
\leq4$, and 0 otherwise. Two normalized incompressible fields are then
constructed as
\begin{equation}
{\bf f}^{(1)}= \frac{\nabla \times {\bf v}^{(1)}}{\left< |\nabla
    \times {\bf v}^{(1)}|^2 \right>^{1/2}},
  \hspace{.5cm} {\bf f}^{(2)}= \frac{\nabla \times {\bf v}^{(2)}}
    {\left< |\nabla \times {\bf v}^{(2)}|^2 \right>^{1/2}}.
\end{equation}
We introduce a correlation between these fields making use of the auxiliary quantity as
\begin{equation}
{\bf \omega}_f = \nabla \times 
    [\sin(\alpha) {\bf f}^{(1)}+\cos(\alpha) {\bf f}^{(2)}]
\end{equation}
where $\alpha$ can take any value between 0 and $\pi/4$. Finally, the
forcing function ${\bf f}$ is given by
\begin{equation}
{\bf f}({\bf k}) = f_0 \left[ \cos(\alpha) {\bf f}^{(1)}({\bf k}) +
    \sin(\alpha) {\bf f}^{(2)} ({\bf k}) + \frac{{\bf \omega}_f({\bf
        k})}{k} \right] ,
\end{equation}
with $f_0$ the forcing amplitude. Note that $\alpha$ controls how much
correlated is ${\bf f}$ with its curl (in fact, this correlation is
proportional to $\sin 2\alpha$). Thus, using $\alpha=0$ gives a random 
forcing function that does not inject helicity on the average. This
allows us to study cases in which there is no inverse cascade (as the
presence of helicity in the three-dimensional flow can result in the
growth of large scale structures as a result of the inverse cascade of
magnetic helicity \citet{Mininni2011}), which will be important to
understand the role of the growth of correlation lengths as the flow
becomes two-dimensional in particle acceleration for large values of
$B_0$.

When a stationary turbulent state was reached and a broad range of 
scales were excited, the test particles were injected and then both
field and particles were simultaniously evolved (for frozen-in-time
simulations, one snapshot of the fields was extracted, and only test
particles were evolved). In the turbulent regime the fluid contains
energy from the outer scale $L$ to the Kolmogorov dissipation scale 
$l_d=(\nu^3/\epsilon_d)^{1/4}$, where $\epsilon_d$ is the average rate
of energy dissipation; then we can define the Kolmogorov dissipation
wavenumber as $k_d=2\pi/l_d$.

We must now introduce the equations for the
test particles, and associate the particle parameters with the
relevant flow parameters.

Test particle dynamics is described by the nonrelativistic equation of
motion:
\begin{equation}
  \frac{d\textbf{v}}{dt} = \alpha(\textbf{E} + \textbf{v} \times
  \textbf{B}), \ \ \ \  \frac{d\textbf{r}}{dt} = \textbf{v}.
\end{equation}
The electric field \textbf{E} can be obtained from the
generalized Ohm's law which can be dimensional scaling by $E_0= v_0 B_0/c$ as follows:
\begin{equation} 
 \textbf{E} =  -\textbf{u}  \times \textbf{B} +
 {\frac{\epsilon}{\rho}}\textbf{J}  \times \textbf{B} - \epsilon
 \nabla p_e + \frac{\textbf{J}}{R_m}.
\end{equation}
The dimensionless parameter $\alpha$ relates particles and MHD field
parameters:
\begin{equation}
\alpha=Z\frac{m_p}{m}\frac{L}{\rho_{ii}},
\end{equation}
where $\rho_{ii}$ is the proton inertial length given by
$\rho_{ii}=m_pc/(e\sqrt{4\pi\rho_0})$, $m$ is the mass of the
particle, $m_p$ is the mass of the proton, and $Z$ is the atomic
number (we will consider $m=m_p$ and $Z=1$). The inverse $1/\alpha$
represents the nominal gyroradius, in units of $L$ and with velocity
$v_0$, and measures the range of scales involved in the system (from
the outer scale of turbulence to the particle gyroradius). One could
expect a value $\alpha > 10^{4}$ specially for space physics and
astrophysical plasmas, which represent a huge computational challenge
due to numerical limitations.

Once the turbulent state was reached, $10000$ test particles were 
randomly distributed in the computational box and the equation of
motion for particles  and the MHD electromagnetic fields were evolved.
Equation (9) for each particle was evolved in time using a
fourth-order Runge-Kutta method, and cubic splines were used to
extrapolate the values of the terms in Eq.~(10) from the
three-dimensional MHD grid to the position of the particles. Particles
were initialized with a Gaussian velocity distribution function, with a  
root mean square (r.m.s.) value of the order of the Alfven
velocity. It is well known that the particle gyroradius has a
significant influence on the particle acceleration mechanisms, and our
aim in this paper is to explore the dynamical effect of MHD turbulence
on acceleration of large gyroradius particles, of the order of the
turbulent dissipation length. Thus, we set $\rho_{ii} = l_d$. We will
loosely call these particles ``protons'' as their gyroradius is at the
end of the inertial range of our MHD simulations(\citet{PD1}).

The second and the third terms in Eq.~(10) are the Hall effect
and the electron pressure gradient respectively. The dimensionless
coefficient $\epsilon$ is the Hall parameter:
\begin{equation} 
 \epsilon = \frac{\rho_{ii}}{L}
\end{equation}

Those two terms in Eq. (10) are important at small scales, especially
at the proton gyroradius but they make
small contributions to the fluid equations through the curl of \textbf{E} (i.e. single-fluid
compressible MHD is still appropriate at large scales).
However one must expect them to impact the acceleration of particles 
by their contribution to the electric field \textbf{E}(\citet{Kulsrud}).

The Hall parameter relates the ion inertial length scale with the energy
containing scale. Thus, for consistency with the test particle
definition (see Eq.~11), we set the value of the Hall parameter
$\epsilon=1/\alpha$ in our simulations. In the MHD description it is
assumed that plasma protons and electrons are in thermal equilibrium,
i.e., their pressures are $p_e=p_i$. Then $p_e = p/2$ with 
$p=p_e+p_i$ the total pressure.  For particles with 
$\alpha = 1/\epsilon$ and $\rho_{ii} = l_d$ (as is the case studied
here), we showed in \citet{Gonzalez2016_1} that the second and
third terms on the r.h.s.~of equation.~(10) give a negligible contribution
to the acceleration. We will however preserve these terms for
consistency. The case of small particle gyroradii (such as particle with a real mass electrons), which
was also previusly reported, those terms become really relevant, and we
will left for a future study about the effect of dynamicall electromagentic fields over that kind
of particles.

\section{\label{sec:level3}RESULTS:}

\subsection{Dynamic vs.~static MHD fields}

\begin{table}
\renewcommand{\thetable}{\arabic{table}}
\caption{Parameters of the simulations discussed in Sec.~III.A. $B_0$
  is the amplitude of the guide field, $M$ is the sonic Mach number,
  $\beta$ is the plasma ``beta,'', $\langle L \rangle$ is the average energy containing
  scale for the flow during the steady-state period, $\langle k_d \rangle$ is the average Kolmogorov dissipation wavenumber, and $\alpha$ is the
  nominal particle gyroradius.}
\label{tab1}
\begin{tabular}{ccccccc}
\toprule
Run \ & \ $B_0$ \  & \ $M$ \ & \ $\beta$ \ & \ \ $\langle L \rangle$ \ & \ $\langle k_d \rangle$  \ & \
       $\alpha$ \\
\colrule
1 \ & \ 0 \ & \ 0.25 \ & \ 6.12 \ & \ 2.31 \ & \ 92 \ & \ 40 \\
2 \ & \ 2 \ & \ 0.25 \ & \ 2.79 \ & \ 1.76 \ & \ 96 \ & \ 52.63 \\
\botrule
\end{tabular}
\end{table}

In this section we compare the particle behavior for static and
dynamic MHD fields, in two different scenarios with and without an
external mean magnetic field.  Table 1. presents the parameters of the
flow and of the particles for the simulations in this section.

\begin{figure}
\begin{center}
{\includegraphics[width =0.8\linewidth]{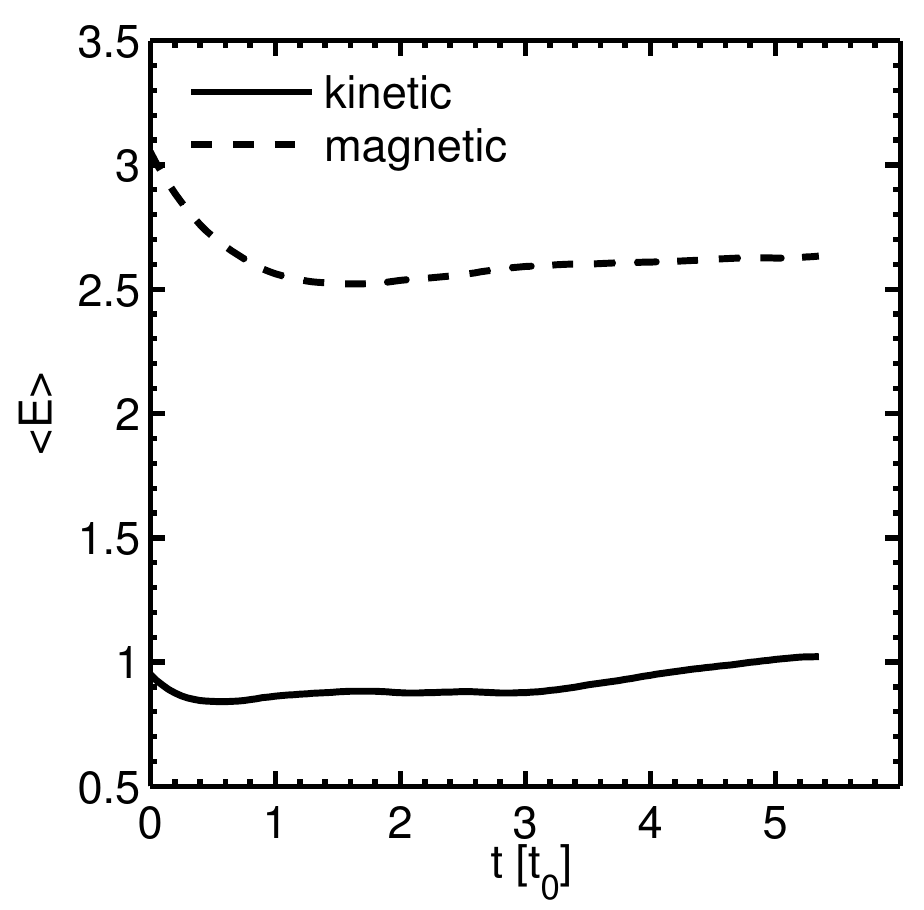}}
{\includegraphics[width =0.8\linewidth]{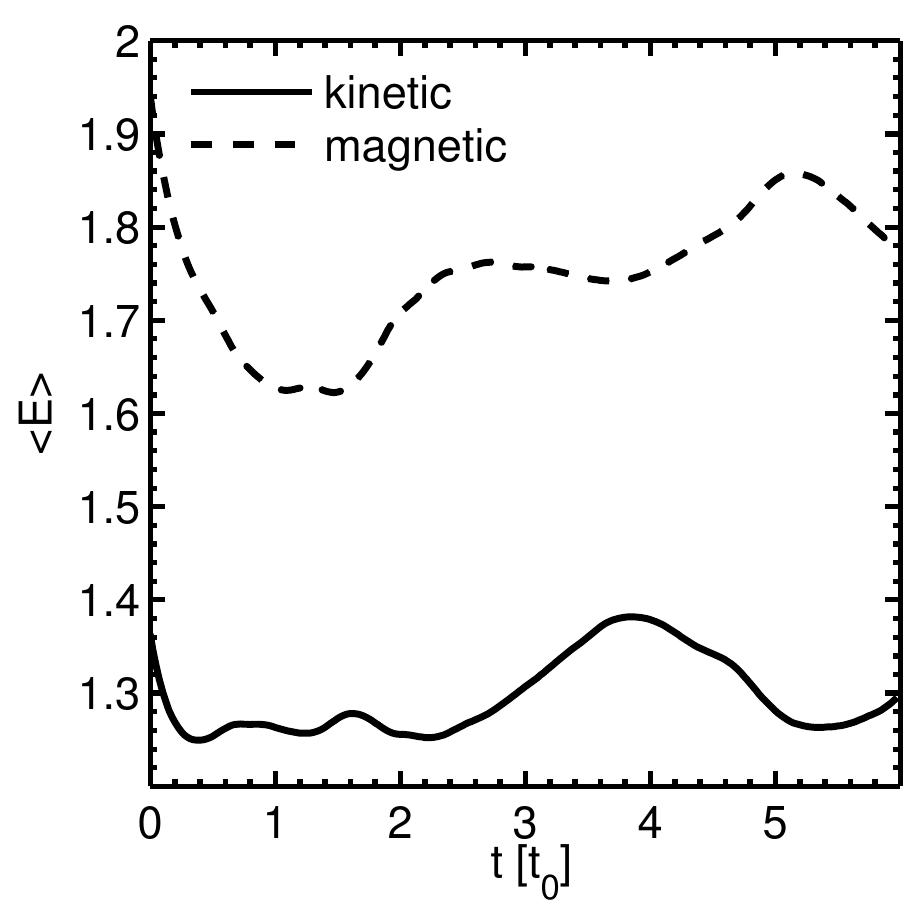}}
\caption{Mean energy of MHD variables as function of time, {\it
    (top)} for a simulation with no mean magnetic field 
mean field simulation ($B_0=0$) and {\it (Bottom)} for simulation with $B_0=2$.}
\end{center}
\end{figure}

In Figure.~1 we show the kinetic and magnetic energy for the
time-evolving simulations with $B_0=0$ and $B_0=2$ during the time that particles are
running simultaniously with the fields.  
The MHD fields are initially started from a null initial state and then 
a forcing is applied from the beginning
of the simulation to obtain a quasi-stationary state (at about $ \sim 15t_0$). Note that the system is in a turbulent steady state, with the energy fluctuating around a mean
value. As a reference, once the particles are added to this flow, it
takes about $5$ turnover times for the particles to cross the entire
simulation box (periodic boundary conditions are used for both the
fluid and the particles, and thus the most energetic particles can
re-enter the box and travel larger distances).

\begin{figure}
\begin{center}
\includegraphics[width = 0.8\linewidth]{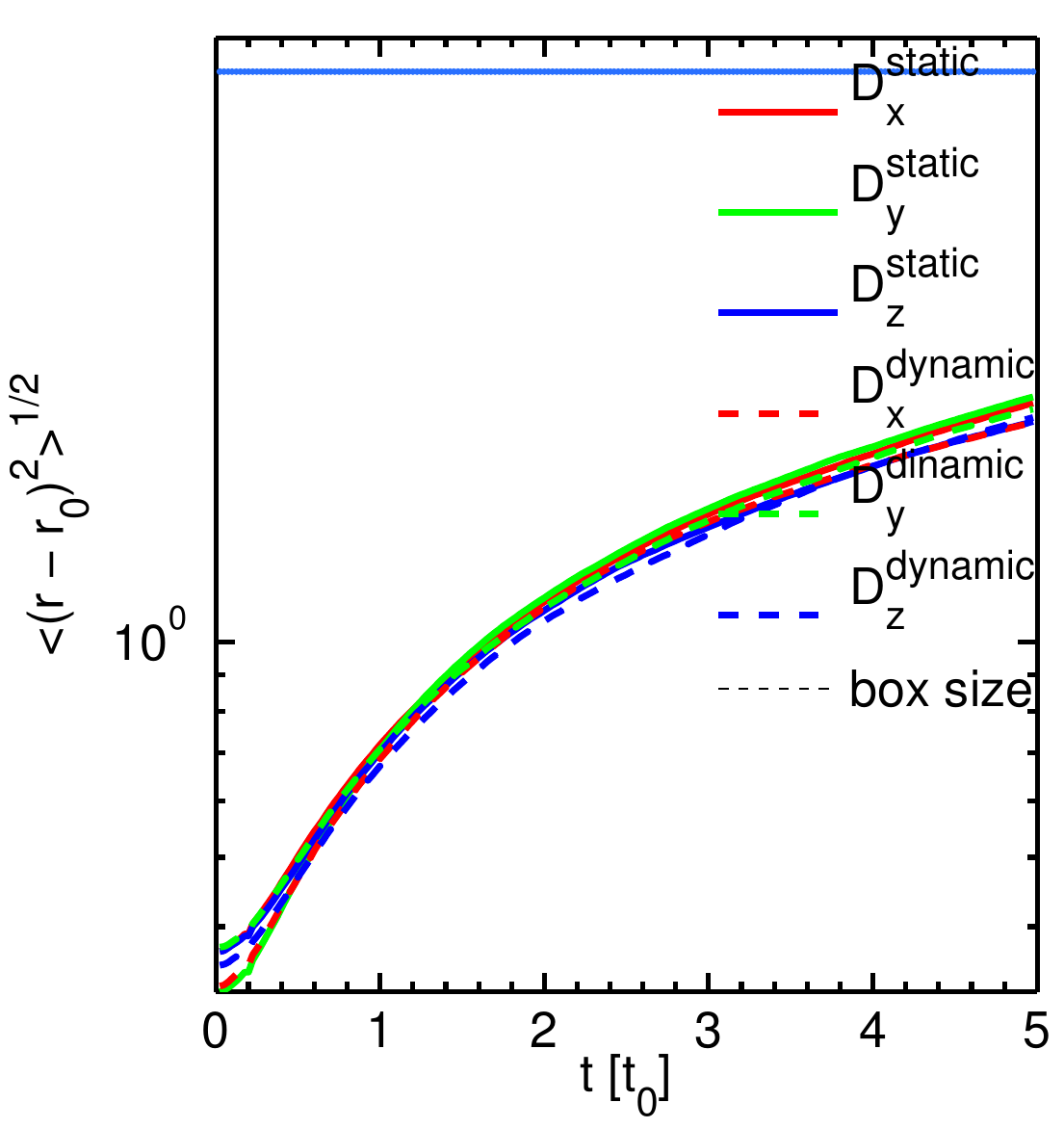}
\includegraphics[width = 0.8\linewidth]{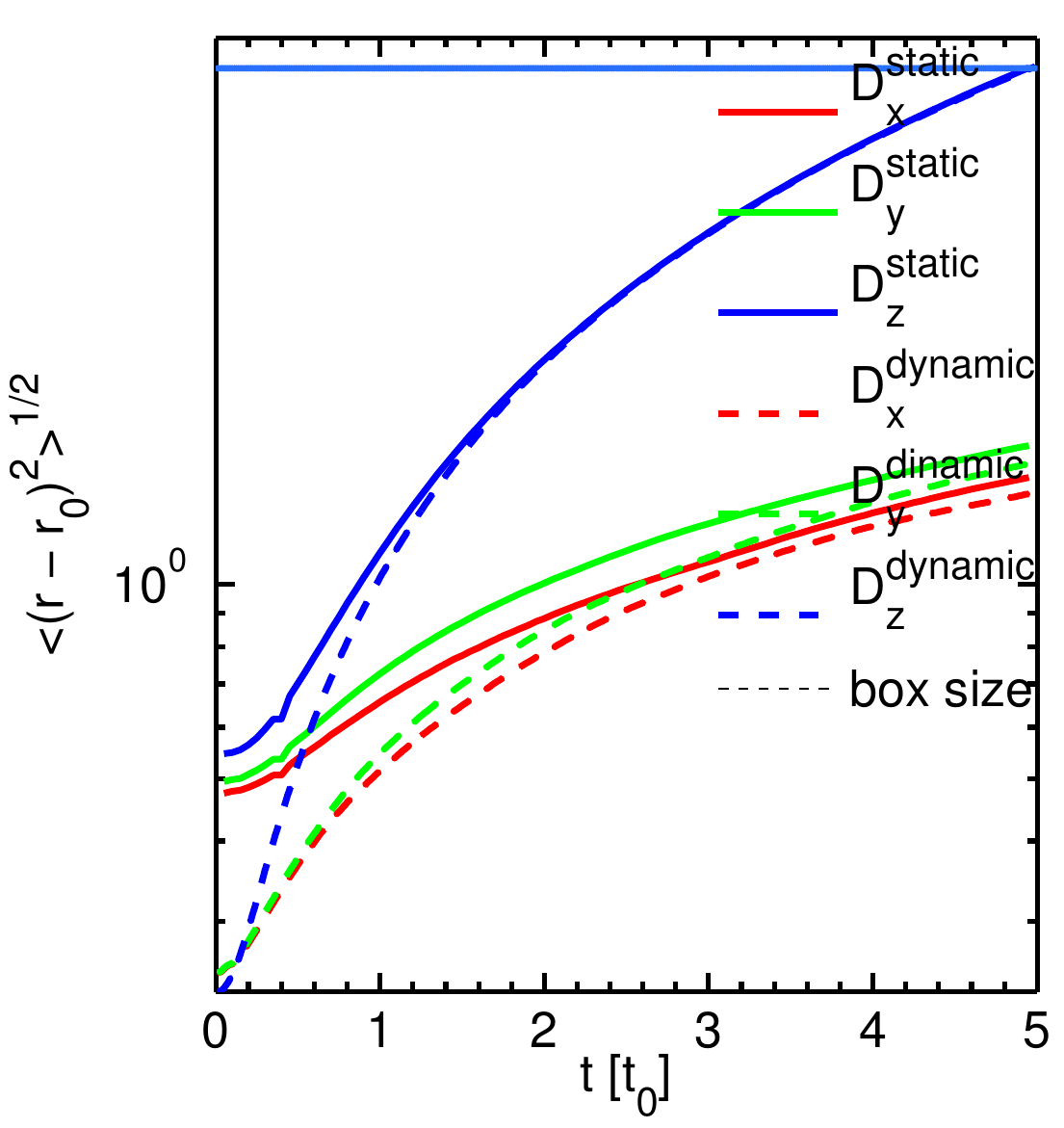}
\caption{Mean square displacement of protons as function of time in
  box units, for the static (solid line) and dynamic (dashed line)
  simulations. {\it{(Top)}} Simulation without mean magnetic field
  ($B_0=0$), and {\it{(Bottom)}} simulation with a mean magnetic field
  $B_0=2$. The dashed line indicates the size of the simulation box.}
\end{center}
\label{mean square velocity}
\end{figure}

The mean square desplacement $D_r$ as a function of
time is shown in Figure.~2, for the simulations with $B_0=0$ and with
$B_0=2$, for both the static and dynamic cases. Again by ``static case,'' we
refer to the case in which a snapshot of the fields is frozen in time,
and the particles are evolved in the electric and magnetic fields (as done in previous
studies, see e.g., Refs.~\citet{PD1,hall,Dalena2014}). In the
``dynamic case'' the equations of motion of the particles are evolved
in time together with the MHD equations, and thus the particles are
affected by the time evolution of the structures in the flow. A small
difference is observed between these two cases, with a slightly
larger mean displacement in the static case. Additionally, the
displacement of the test particles in the $z$ direction in the
simulation with $B_0=2$ changes significantly (when compared with the 
case with $B_0=0$), as a result of the test particles following the
direction of the mean magnetic field. While in the run with $B_0=0$
the dispersion is isotropic (i.e., the mean displacement is the same within statistical 
errors for the $x$, $y$, and $z$ directions), in the run with $B_0=2$ much
larger displacements are observed in the $z$ direction.

\begin{figure}
\begin{center}
{\includegraphics[width = 0.8\linewidth]{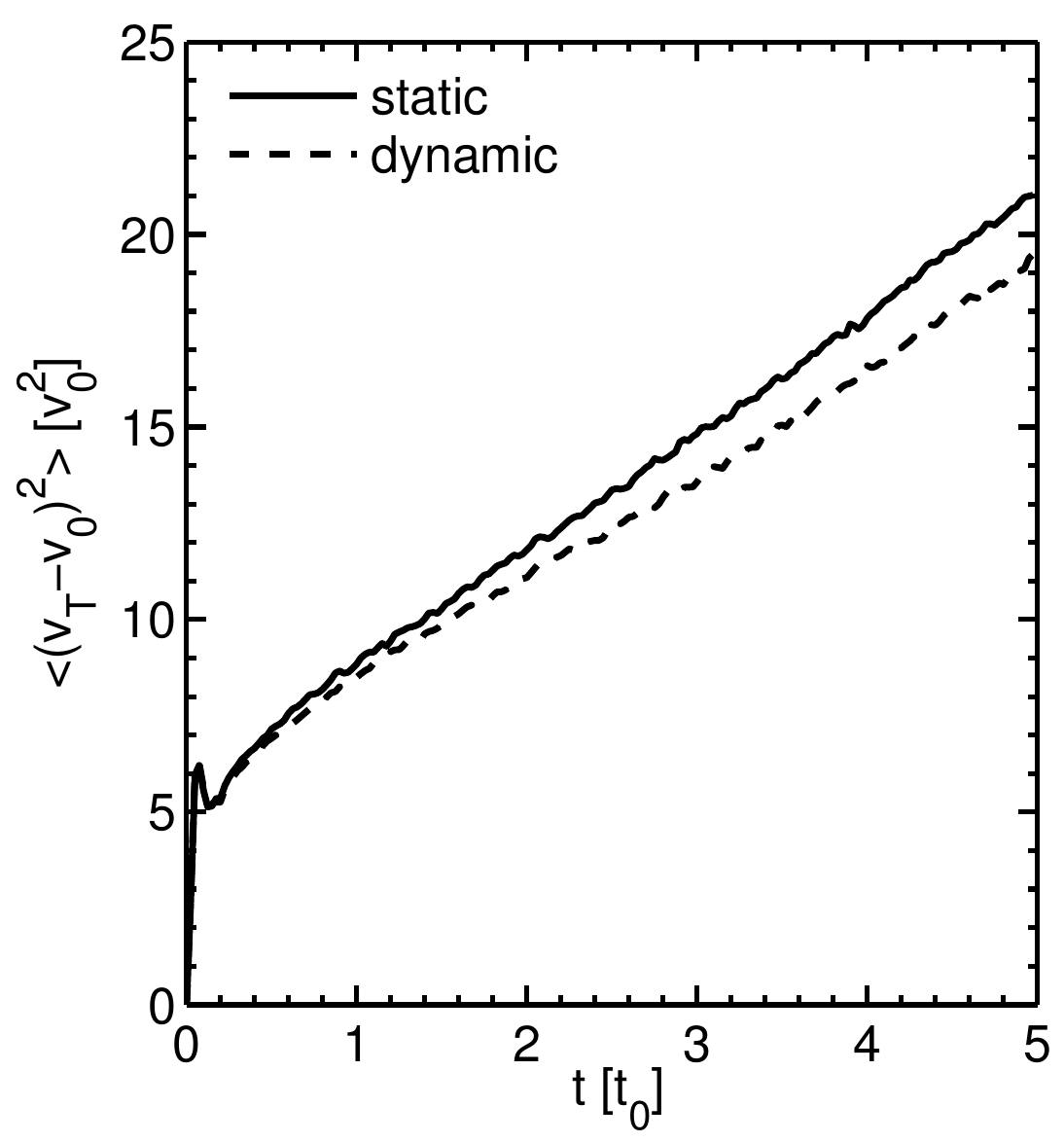}}
{\includegraphics[width = 0.8\linewidth]{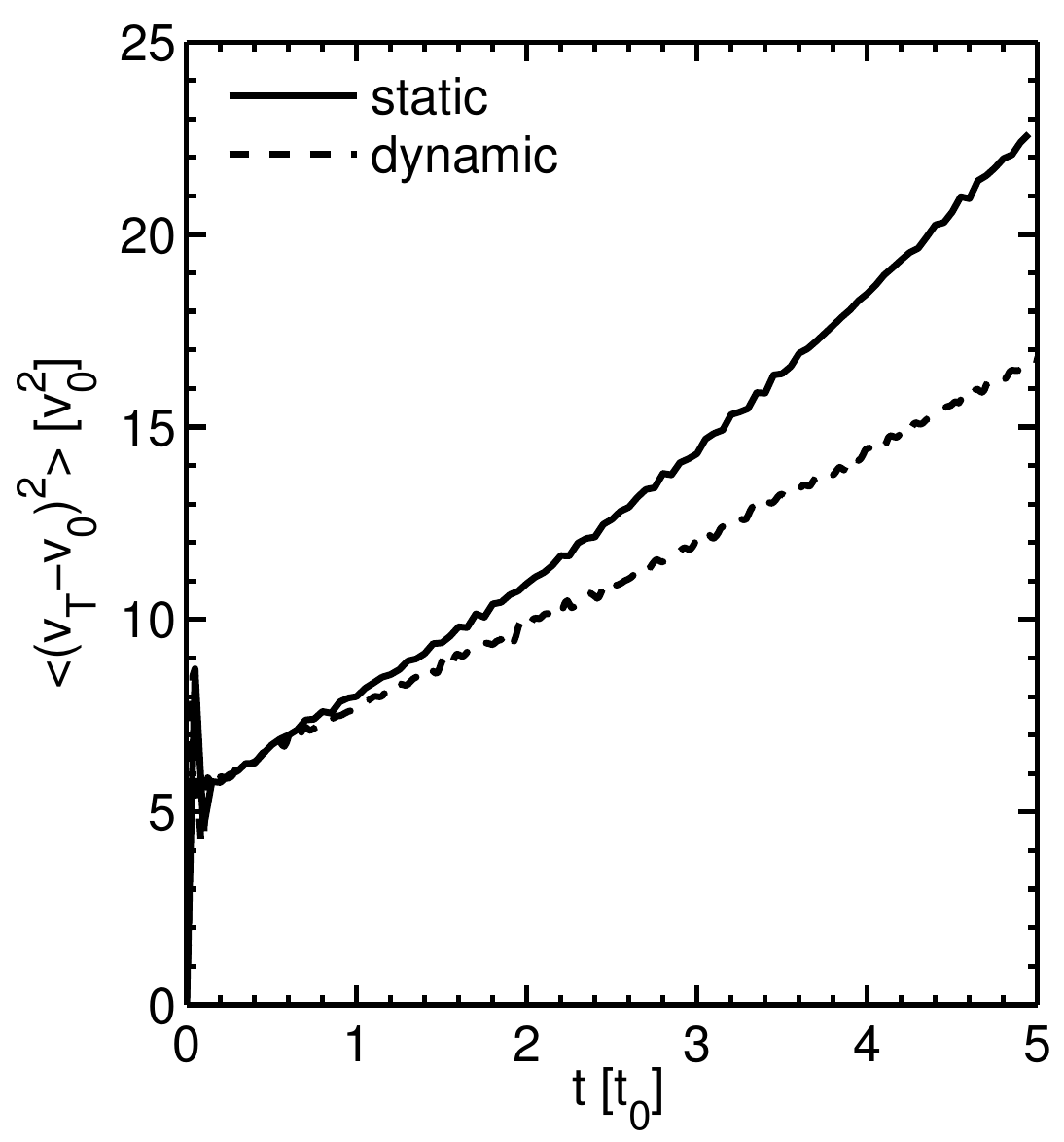}}
\caption{Mean square velocity of protons as function of time for
  the static (solid line) and dynamic (dashed line) cases. {\it (Top)}
  Simulation without mean magnetic field ($B_0=0$) and {\it (Bottom)}
  with mean magnetic field ($B_0=2$).}
\end{center}
\end{figure}

In Figure.~3 we show the mean square velocity  as function of time,
$v_T^2 = v_\perp^2+v_\parallel^2$, i.e., the sum of the squared
parallel and perpendicular components of the test particle velocities,
averaged over all test particles. Results for the isotropic and the
anisotropic cases are shown, comparing also the static and dynamic
cases. A smaller mean velocity at late times is observed in the
dynamic case, specially for the case with $B_0=2$.

The progessive increment in the mean square velocity is the result 
of the interaction with the electric fields near the current sheets found along
the particle trajectory, as was described in a previous paper
\citet{Gonzalez2016_1}. In the isotropic case, the structures are
randomly distributed in the box and there is no privileged direction;
as a result the current sheets are not aligned and are instead
randomly oriented. This situation and the lack of a large magnetic field make
the particles interact with several different structures along its path and to be exposed
to the strong gradient electric field regions for long times.  The reduced final r.m.s velocity in the
dynamic case, instead, is associated with the dynamic coherent structures that trap particles and
reduces this exposure time that is an essential component for particle acceleration.

\begin{figure}
\begin{center}
{\includegraphics[width = 0.8\linewidth]{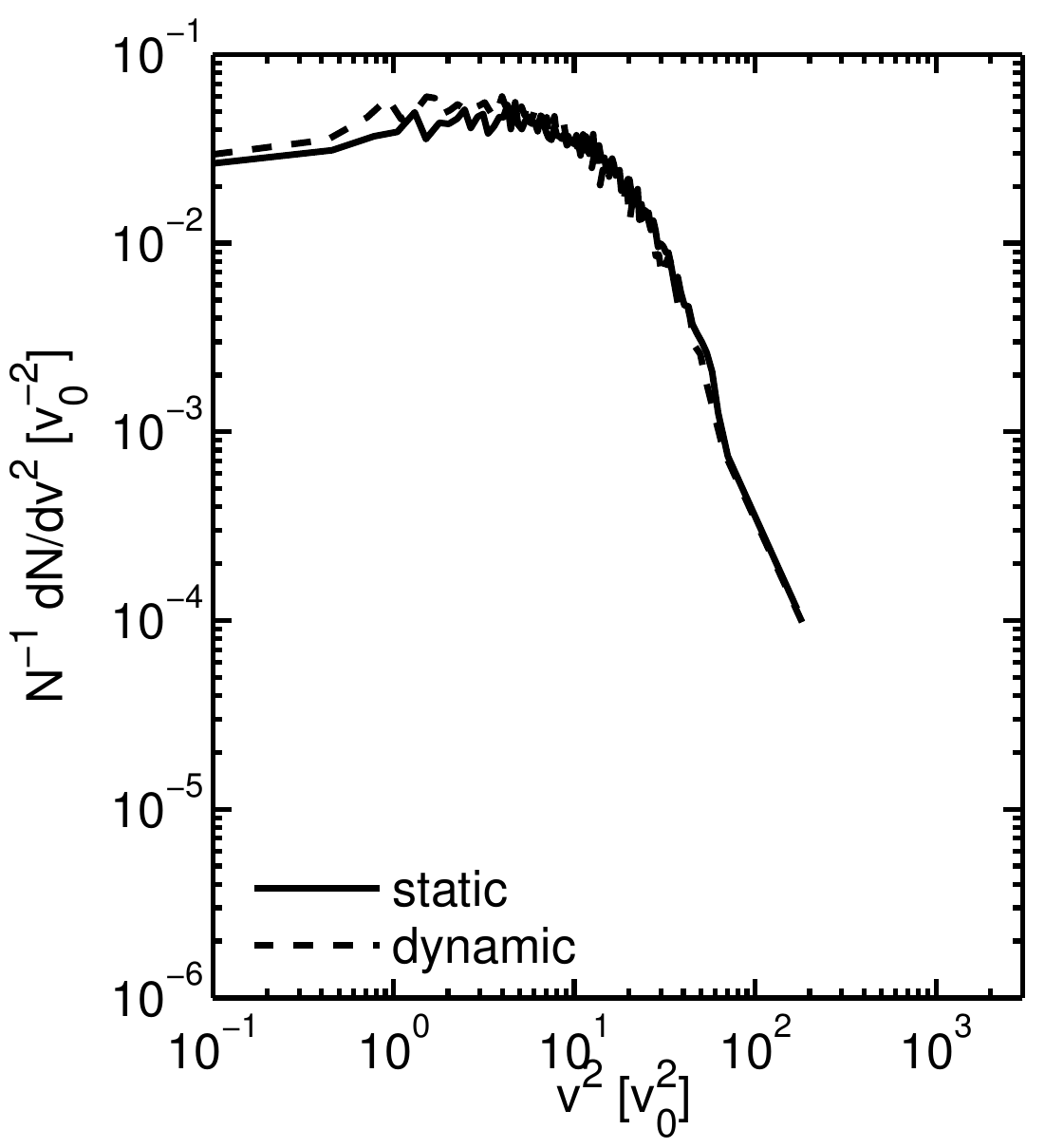}}
{\includegraphics[width = 0.8\linewidth]{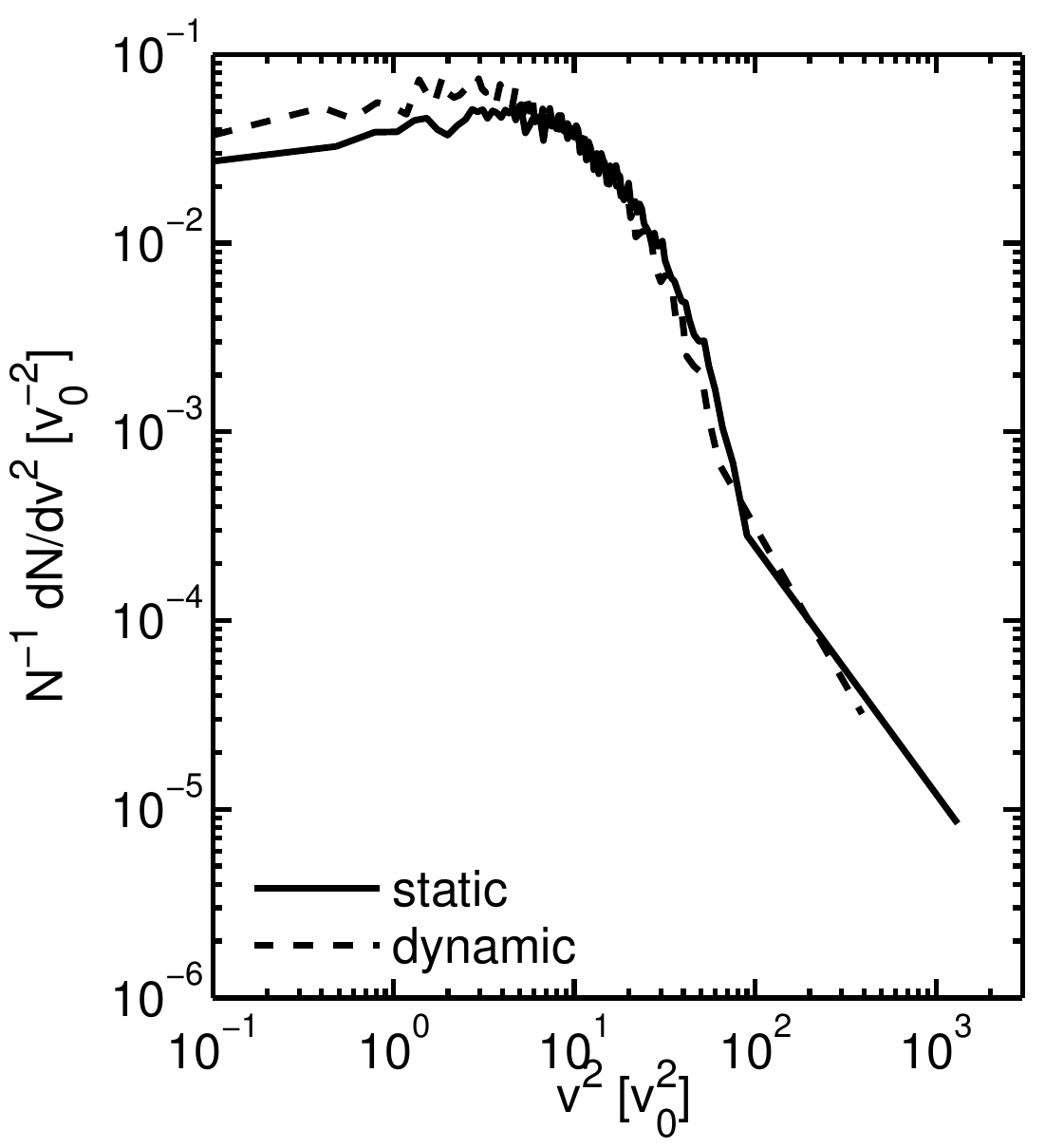}}
\caption{Probability density function of proton energies for the 
static and dynamic cases: {\it (Top)} Simulation without mean magnetic field
($B_0=0$) and {\it (Bottom)} simulation with mean magnetic field ($B_0=2$).}
\end{center}
\label{mean square velocity}
\end{figure}

In Figure.~4 we show the probability density funtion (PDF) of 
particle energy for the isotropic and anisotropic cases (and for both
the static and dynamic cases). In the isotropic case, the distributions
for static and dynamic fields are very similar. There is only a
slightly larger value of probability in the core of the distribution
(i.e., for low squared velocities) in the dynamic case, which can be
understood as we already noted that in the dynamic case the
acceleration of particles is slightly less efficient. However, in the
anisotropic case these differences become somewhat larger (although still small 
in absolute terms), accompanied 
with a shorter tail in the PDF for the dynamic case at large squared
velocities. In the presence of the the mean field more particles 
are accelerated, and with larger energies, in the static case than in
the dynamic case.

As discussed above, visual exploration of the particles trajectories
indicate that the fact that particles are more accelerated in the
static case than in the dynamic case is the result of particles being
able to interact for longer times with multiple current sheets (i.e.,
they can move from one current sheet to another, and cross strong electric
fields gradients in their paths), and thus can obtain larger
energies. Meanwhile, in the dynamic case the particles are trapped for
longer times in an individual current sheet advected by the flow, and
cannot easily move in the strong gradient regions, which is the
most efficient process to accelerate them.

With this in mind, in the next section we explore the effect of the
mean magnetic field  on particle acceleration in the dynamically
evolving turbulent electromagentic field. To this end, we analize
particle energization varying the guide field from zero up to a strong
case with with $B_0=8$.

\subsection{Effect of $B_0$}

\begin{table}
\renewcommand{\thetable}{\arabic{table}}
\caption{Parameters of the simulations with varying $B_0$, discussed
  in Sec. III.B. $B_0$ is the amplitude of the guide field, $\langle
  v^2\rangle$ and $\langle b^2 \rangle$ are respectively the mean
  kinetic and magnetic energies in the turbulent steady state (for a
  fluid with mass density $\rho = 1$), $\beta$ is the plasma ``beta,'', $\langle L \rangle$ is the average energy containing
  scale for the flow during the steady-state period, $\langle k_d \rangle$ is the average Kolmogorov dissipation wavenumber, and $\alpha$ is the
  nominal particle gyroradius.}
\begin{tabular}{ccccccccc}
\toprule
$B_0$ \ & \ $\langle v^2\rangle$ \ & \ $\langle b^2 \rangle$ \
  & \ $\beta$ \ & \ $\langle L \rangle$ \ & \ $\langle k_d \rangle$  \ & \ $\alpha$ \  \\
\colrule
 0 \ & \ 0.91 \ & \ 2.61 \ & \ 6.12 \ & \ 2.41 \ & 
    \ 92 \ & \ 40 \ \\
 2 \ & \ 1.29 \ & \ 1.74  \ & \ 2.79 \ & \ 1.78 \ & 
    \ 96 \ & \ 52.63 \ \\
 4 \ & \ 1.12 \ & \ 1.28  \ & \ 0.93 \ & \ 2.28 \ & 
    \ 86 \ & \ 40 \ \\
 8 \ & \ 1.22 \ & \ 1.13  & \ 0.24 \ & \ 2.28 \ & 
    \ 92.3 \ & \ 43.48 \ \\
\botrule
\end{tabular}
\label{tab2}
\end{table}

\begin{figure}
\begin{center}
\includegraphics[width = 0.8\linewidth]{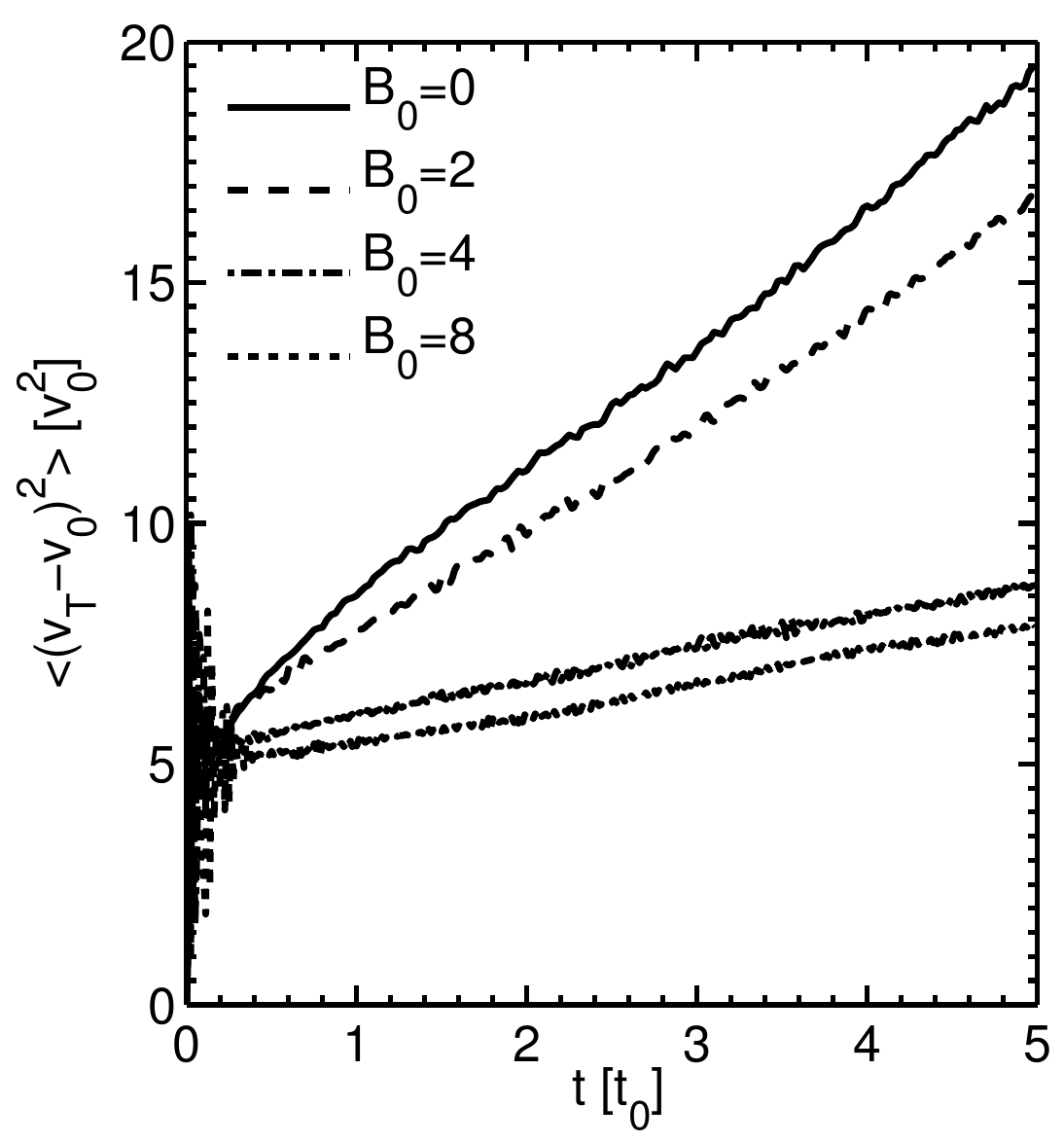}
\caption{Mean square total velocity of protons as a function of time
  for different simulations, changing the mean magnetic field value
  $B_0$.}
\end{center}
\label{mean square velocity}
\end{figure}

The aim of this section is to show the effect of the mean magnetic
field $B_0$ on the flow features, and on the resulting particle
acceleration. It is well known that MHD flows with an imposed strong
magnetic field suffer a transition form a three-dimensional (3D) state
towards a two-dimensional (2D) state (\citet{Alexakis2011,sujovolski2016}). The 
relevance of this anisotropy
has been discussed by many authors in the recent years, specially in the
context of the solar wind problem. This transition from a 3D to a 2D
state is accompanied by the transfer of energy towards modes with
small parallel wavenumber  (i.e., by the increase of the correlation
length of the structures in the direction of the mean field \citet{Alexakis2011}), and and is associated with the development of the
conditions that would establish an inverse cascade of the squared vector 
potential if the flow becomes 2D (\citet{Mininni2005,wareing2010}). Both effects (although in different
ways) result in the growth in size of the structures in the flow. We
thus now look for the effect of the resulting anisotropic field and
increased correlation lengths on test particle acceleration. In Table
2. we show the parameters of the simulations presented in this section.
In the Table it can be noted that the plasma beta is not equal to 1 specially 
for the
isotropic case. It is important to remark that the proton gyroradius and the 
proton inertial length are not rigorously equal in those simulations.

In Figure.~5 we show the r.m.s velocity as a function of time for
all the simulations with different values of the mean magnetic
field. It is observed that as the mean field increases, the particles
velocity at the end of the simulation is reduced, and the acceleration
process is thus diminished. The reduction of the r.m.s velocity with
increasing mean magnetic field is an effect of the growth in the size
of the structures; as current sheets become elongated (in the direction of
the mean field resulting from the anisotropy), and wider (in the
direction perpendicular to the mean field resulting from the flow
two-dimensionalization), the particles get trapped in the structures
(i.e. the particles are magnetized)
and cannot easily be exposed for longer times
to the accelerating gradient electric fields in the vicinity of current sheets.

\begin{figure}
\begin{center}
\includegraphics[width = 0.8\linewidth]{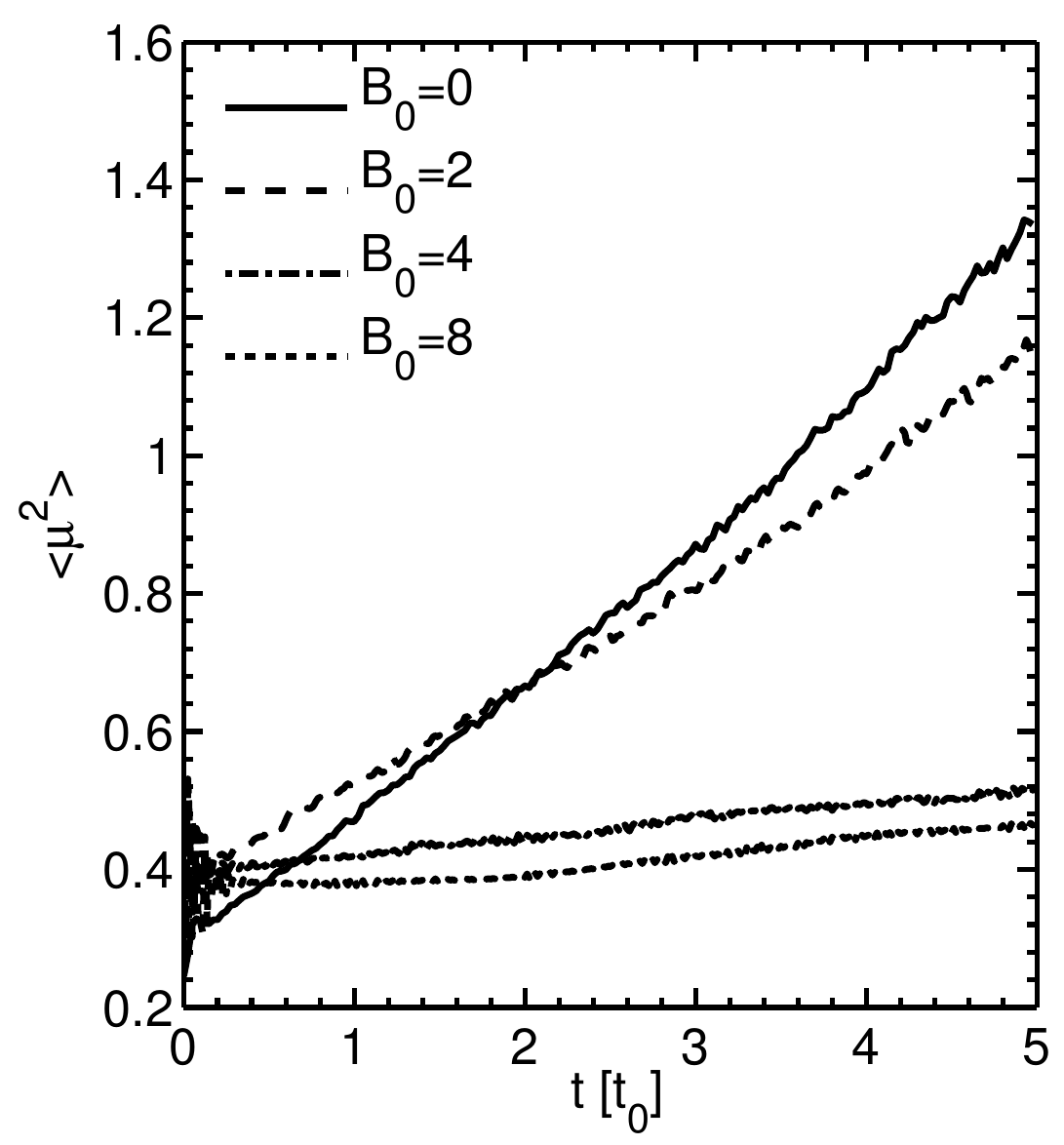}
\caption{Mean square magnetic moment of protons as a function of time
  for different simulations, changing the mean magnetic field value
  $B_0$.}
\end{center}
\label{mean square velocity}
\end{figure}

As observed before, in the isotropic case ($B_0=0$) particles find
structures distributed in all directions, and thus the final velocity
is greater than in all the other cases with non-zero mean magnetic
field. 
This is because the perpendicular energy gained by the the particle is
decreased more substantially in presence of a strong mean 
magnetic field, confirming the argument
about the effect of the anisotropy mentioned above.

In Figure.~6 we show the mean square magnetic moment for protons as funtion of time for 
all the simulations with different values of the mean magentic field. The magnetic 
moment $\mu=W_{\perp}/2B$, with $W_{\perp}$
the perpendicular energy of the particle, is one of the
adiabatic invariant of charged particles dynamic in magnetic fields. it has  
important consequences in the dynamic of particles and determine how magnetized is a 
particle, that is how much 
attached is a particle to a magnetic field line. Variation of magnetic moment changes with turbulence 
parameters was studied in some detail by \citet{Dalena2012}. It is observed that 
the mean square magnetic moment of protons increase its variation in time
as the mean magnetic field is decreased.
This means that protons become more demagnetized for lower values of the 
mean magnetic field. 
On the other hand, as 
the mean magnetic field is increased 
it can be observed an
onset of magnetic moment conservation.  
This means that particles become magnetized (attached to the field
lines).
Then, in those cases, particles propagate 
along the mean field direction and the perpendicular crossing
through different current sheet structures become supressed.

\begin{figure}
\begin{center}
\includegraphics[width = 0.8\linewidth]{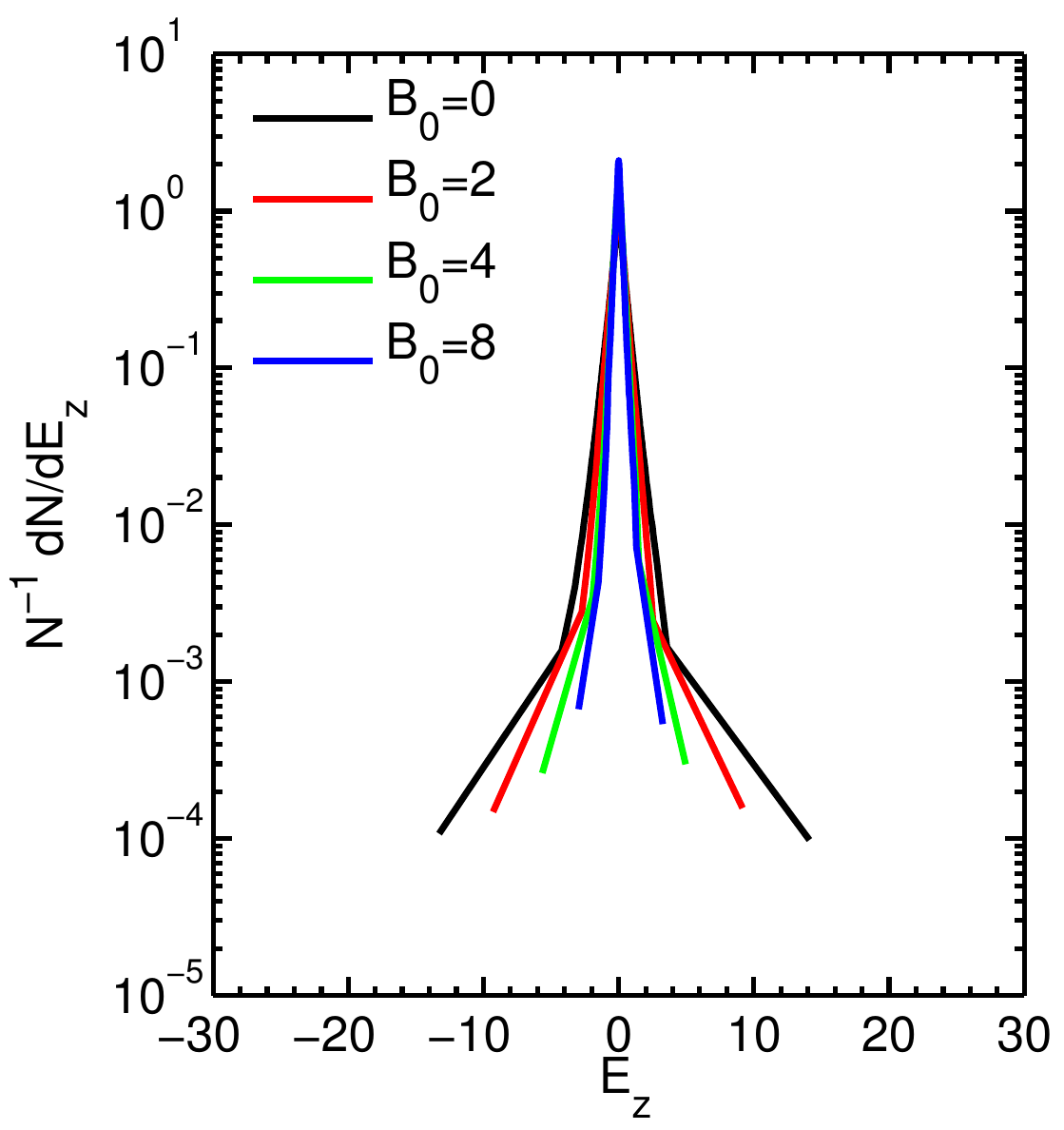}
\includegraphics[width = 0.8\linewidth]{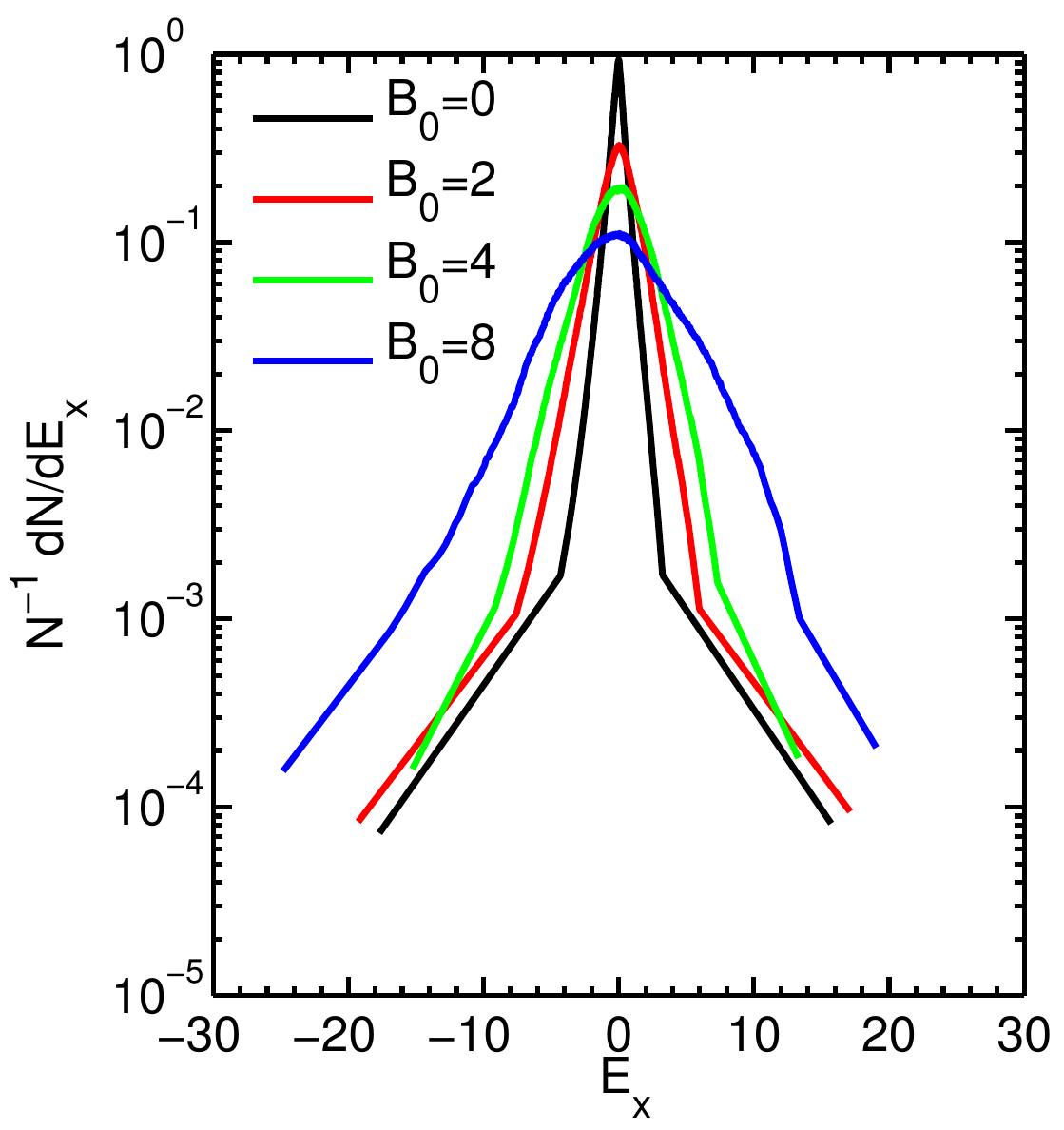}
\caption{Probability distribution function of the electric field in all 
simulations. {\it (Top)}  The parallel ($z$ component) of the electric field $E_z$ and {\it (Bottom)}
  The perpendicular (\textit{x} component) of the electric field $E_x$.}
\end{center}
\label{mean square velocity}
\end{figure}

\begin{figure*}
\begin{center}
\hspace*{-0.5cm}
{\includegraphics[width = 2in]{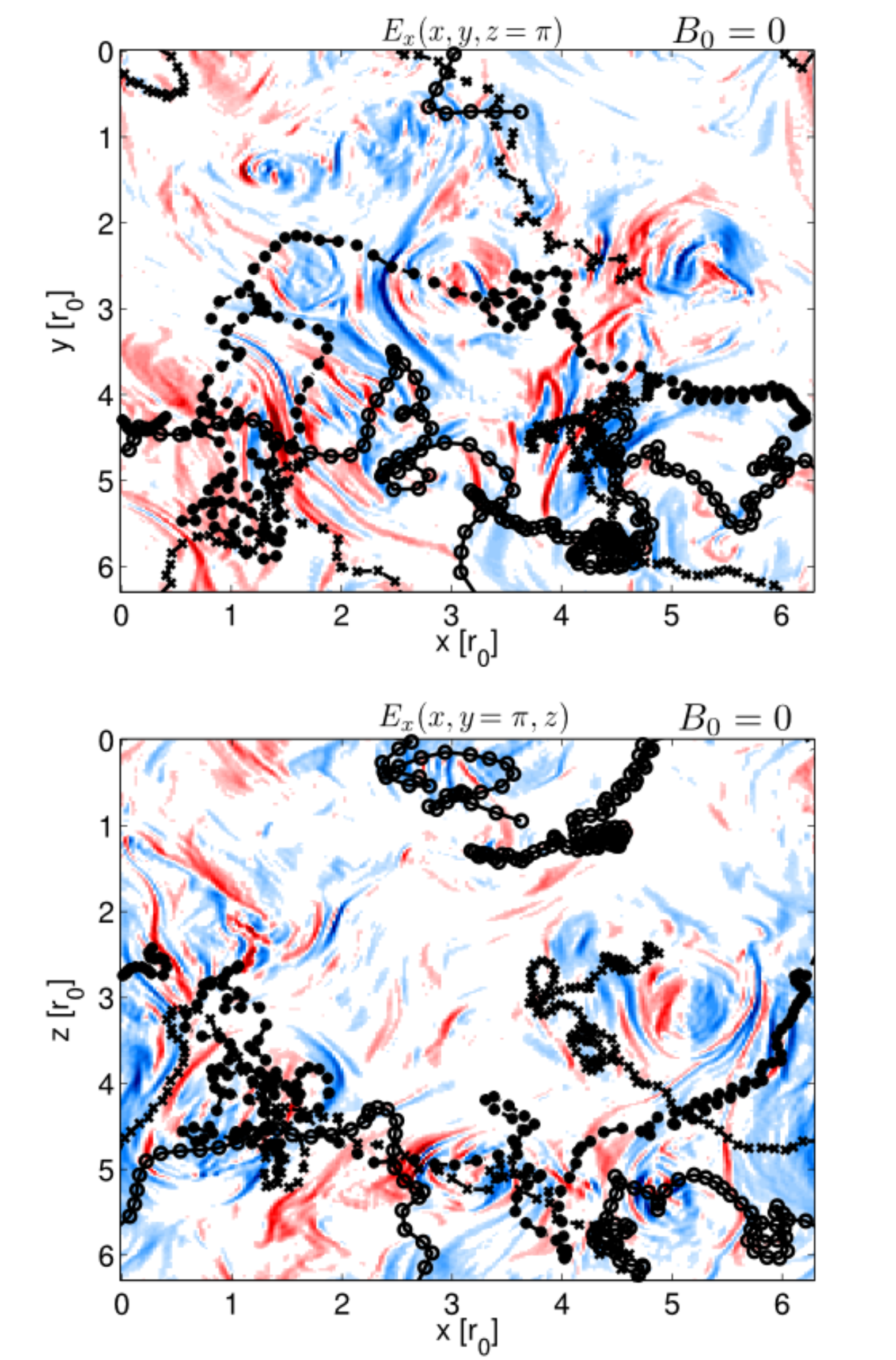}}\hspace*{-0.5cm}
{\includegraphics[width = 2in]{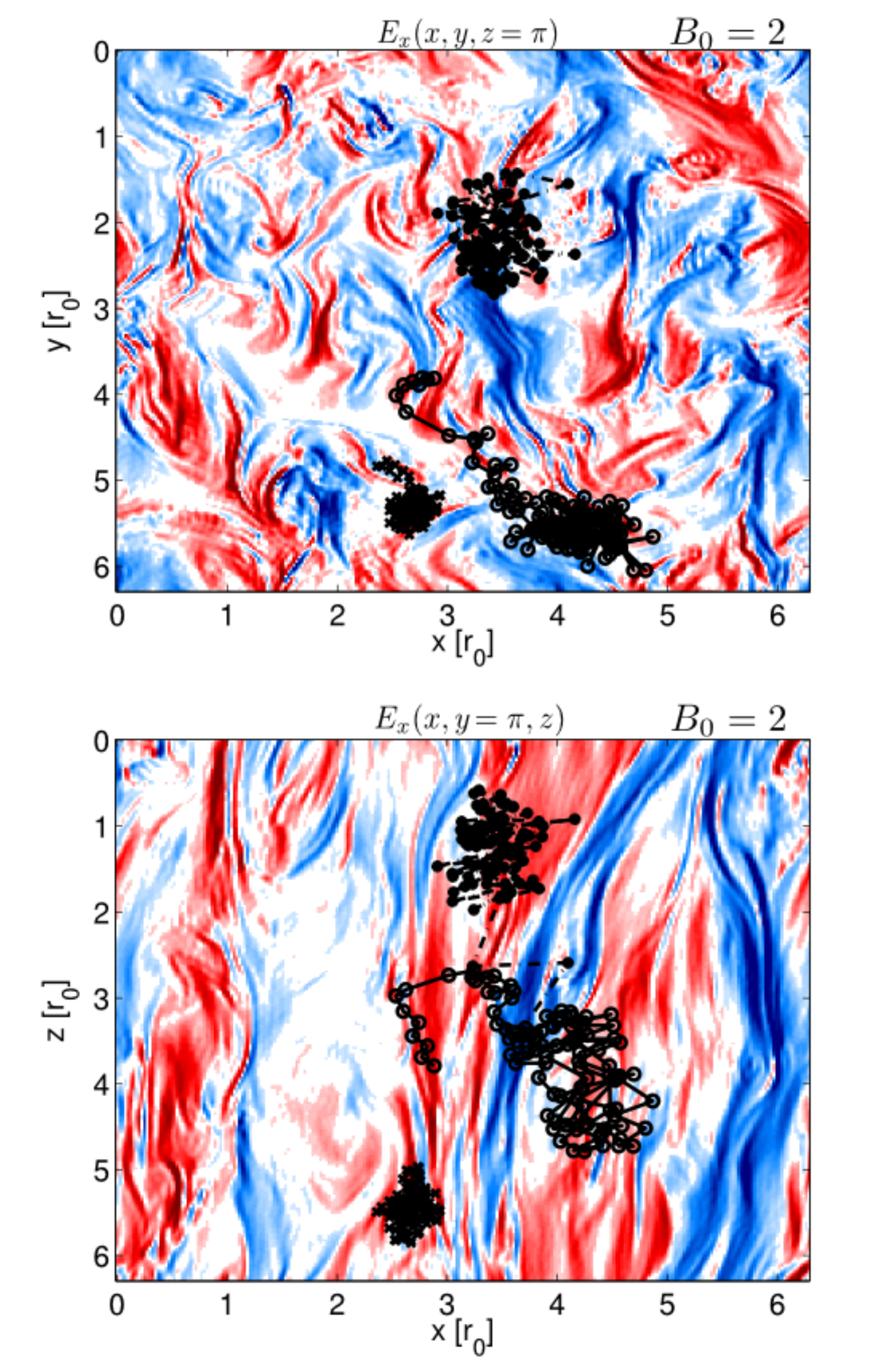}}\hspace*{-0.5cm}
{\includegraphics[width = 2in]{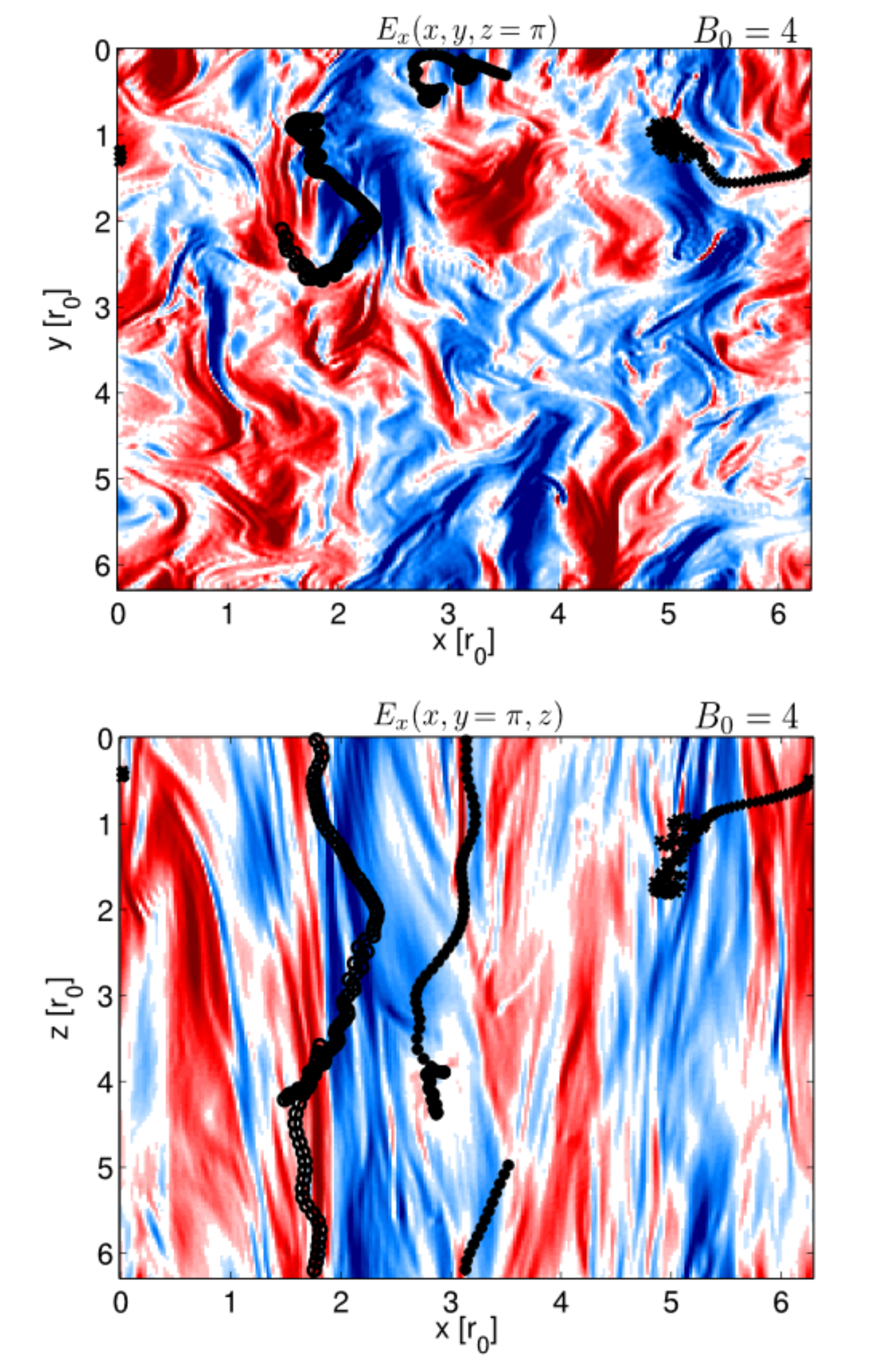}}\hspace*{-0.5cm}
{\includegraphics[width = 2in]{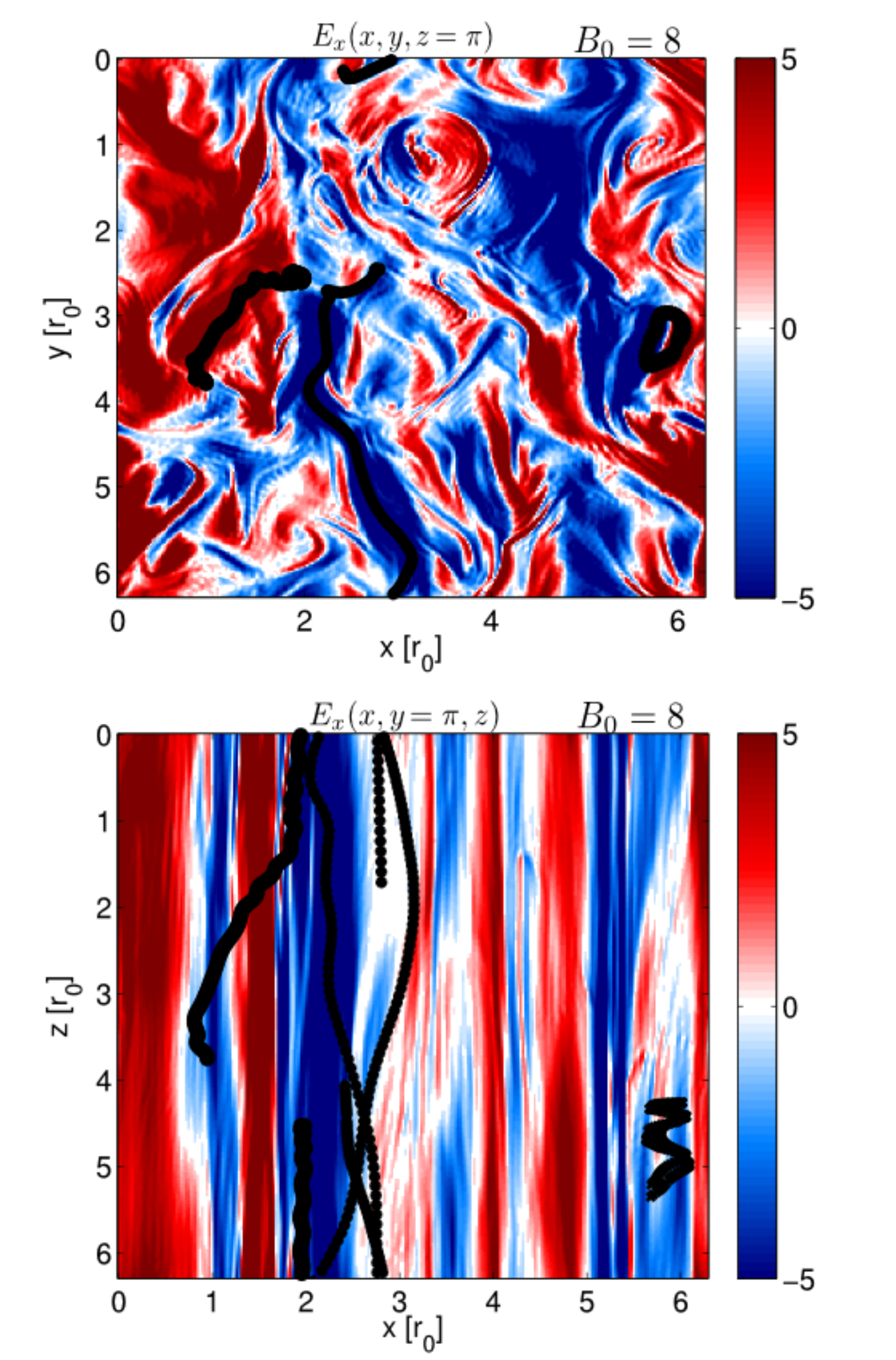}}
\caption{{\it (Top row)} \textit{x-y} cross-section of the perpendicular (\textit{x}
  component) electric field ($E_x$) in the simulation box, for different values of $B_0$. 
 {\it (Bottom row)} \textit{x-z} cross-section of the perpendicular (\textit{x}
  component) electric field ($E_x$). In both rows,
 from left to right, the following simulations are shown:  $B_0=0$, 
  $B_0=2$,  $B_0=4$, and  $B_0= 8$ respectively. The black signs and lines
 show the trajectories of three of the most energetic protons in each
 simulation.}
\end{center}
\label{mean square velocity}
\end{figure*}

The PDFs of the values of the electric field also help to understand
this scenario. The PDFs of the electric field Cartesian components are
shown in Figure.~7. The parallel ($z$ component) of the electric field
takes larger values in the isotropic case than in the anisotropic
cases. As already mentioned, this is because current sheets in the
former case are oriented in all directions, and the electric field is 
larger due to the large values of the velocity field. As $B_0$ is
increased, the the parallel (z) electric field becomes weaker. The
perpendicular (\textit{x} component) of the electric field shows the
opposite behavior: electric field values are larger as the mean
magnetic field increases. This is caused mainly by the first term 
in equation.~(10) (the term containing the mean magnetic field value). 
In spite of this, particle energization
does not follow the same tendency because as noticed the particles cannot leave the
currents structures and then cannot interact with strong electric field 
gradients in regions nearby 
the current sheets interfaces.
That is, what is relevant for perpendicular particle energization is not
the absolute value of the electric field but rather the exposure
time of particles to strong gradients of that electric field.

To confirm this scenario, in Figure.~8 we show the \textit{x-y}
cross-section of the perpendicular component of the electric field
$E_x$, and the \textit{x-z} cross-section of $E_x$ for all the
simulations discussed in this section (from left to right we show the
cases with $B_0=0$ up to the strongest case with $B_0=8$). The
trajectory of three of the most energetic particles are shown as well
with solid lines and marks (using plus, circles and crosses for each
particle).

Note the particle behavior and the importance of the structure size on
particle energization. In the isotropic case it is observed that the
structures are distributed in all directions, and there are no
differences between both cross-sections of the electric field. In
contrast, the cases with nonzero mean magnetic field show the
structures aligned with the mean field as clearly seen in the bottom
panels of Figure.~8. As the mean magnetic field increases, the flow
becomes more anisotropic and for the strongest mean field 
the variations in the $z$ direction are small.

The \textit{x-y} cross-sections give us an idea of the size of the
structures. It is observed that as the mean magnetic field increases,
the width of the structures also increase, and a transition from a
flow with randomly distributed structures with a width of the order of
the dissipation scale can be seen for the isotropic case, moving
towards a flow with wider structures in the cases with a
non-zero mean magnetic field.

The trajectories of the particles show the importance of the structure
size. While trajectories reminiscent of stochastic motion are observed
at the isotropic case, where the particle passes through many
structures finding strong electric field gradient regions along its path, as the mean magnetic field increases
the particle trajectories become more elongated and particles cannot go across structures. For 
the $B_0=4$ and $B_0=8$ cases, the trajectories show
that particles drift across a few structures, almost staying around a
single one and no longer allowing the particle to stay in 
strong electric field gradient regions. This is the reason 
why particles do not gain as much energy as in the isotropic case, as the random hopping from one
structure to the next is the most efficient acceleration mechanism as
observed in \citet{Gonzalez2016_1}.

\section{\label{sec:level4}DISCUSSION:}

In this paper we studied the effect of dynamically evolving MHD
turbulence on test particle acceleration. To this end, we solved
numerically the MHD equations, and solved the equations for test
particles either together with the flow evolution, or in
frozen-in-time fields obtained from snapshots of the electromagnetic
fields in the MHD simulations. This case of ``static MHD turbulence''
has been used before to investigate particle energization phenomena
(\citet{PD1,Dalena2014,Gonzalez2016_1}). We found some differences 
between both models, with a slight reduction of particle acceleration in
the dynamical case. This result was obtained for ``protons,'' or test
particles with gyroradius of the order of the flow dissipation scale,
which is the case we specifically analyzed here. The reduction in the
acceleration rate is due to particle trapping in the dynamic current
structures, that makes more difficult for particles to move through
strong electric field gradient regions located at the interface between structures
whereby particles are accelerated.

The study was motivated by the fact that several of the previous works
done on particle acceleration considered static MHD turbulence. Even
though there are some previous papers which have deal with dynamic   
problem(\citet{CHO1,Weidl2015,Bogdan2014,hussein2016}), most of those studies
consider incompressible flow models. Also, a detailed comparison
between static and dynamic fields and their effect of particle
acceleration was lacking.

This paper is most closely related with \citet{lehe2009heating} due to the compressible
turbulence and dynamic model which both have used. In that paper the authors conclude that the particle
heating is due to the cyclotron resonance with Alfv\'en waves in the system when the waves frecuencies 
are of the order of the particle gyrofrecuency. In this paper we have shown that the perpendicular  
proton heating is almost the same for the dynamic and static cases, which means that we can obtain the
same results with or without allowing waves in the system. In fact, this result is very important 
because it confirm the importance of coherent structures in particle energization. Also, the authors 
discused about the relevance of the results in the context of solar wind and solar corona, and  they 
claim that the results can not be directly applied to those astrophysical systems,
mainly because the 
limitation of numerical resolution and the short scale separation one can obtain using this numerical 
method.  We agree with the authors about the needed to extrapolate this results to a higher numerical 
resolution which would lead to recreate closely the corona and solar wind.

Aditionally, we investigated the effect of anistropy caused by the mean
magnetic field on the resulting particle acceleration, for which it
was observed an important adverse effect on particle energization. The
transition from a three-dimensional towards a two-dimensional MHD
state, which can be also accompanied by an inverse energy cascade for
very strong magnetic guide fields, causes an increase in the structure
sizes, impacting on the reduction of the particle acceleration as test
particles are trapped in wider current channels, and exhibit an almost 
magnetized state with, in average, conservation of magnetic moment.

Considering the results reported in \citet{Gonzalez2016_1}, where
we showed that the flow compressibility affects particle energization,
in this work we made sure that no important variations in the
turbulent Mach number were present betweem different runs. To this
end we also considered forced simulations (instead of freely decaying
ones), to be able to study particle acceleration in a turbulent steady
state with a well defined mean Mach number. To prevent effects
associated with a possible inverse cascade of magnetic helicity, or of
cross-correlations between the velocity and magnetic fields, we
implemented a forcing mechanism that warranties no net injection of
kinetic helicity, magnetic helicity, or cross-helicity.

In the same way, the model we use in this paper for the test particles
includes electron pressure effects and the Hall current in the generalized Ohm's
law for particle motion computation. It is noted that we have not included those terms in the induction equation
based on the assumption that we can neglect them at 
large fluid scales \citet{hall}
While
retaining those effects for particle motion we found
that pressure effects can have an important contribution to the
energization of particles with small gyroradius \citet{Gonzalez2016_1} 
(compared with the
width of the current sheets). For
particles with gyroradius of the order of the MHD dissipation scale
(called ``protons'' here) these effects are not important, with the
main mechanism reponsible for acceleration being the MHD electric
field and its fast changes as the particles move from one current
sheet to another, allowing particles to be exposed to strong electric 
field gradients at the interface of the structures.

The results support and reaffirm the importance of coherent structures
on particle  acceleration, and their possible relevance for the solar
wind. This latter problem has been analyzed using different
approaches, from spacecrafts data measurements(\citet{Tessein2015}) to
different numerical  schemes covering fluid and kinetic
descriptions 
(\citet{AmbrosianoEA88,Greco2008,ServidioKinetic,karimabadi2013,DrakeEA06}).
The coherent structures in this case appear by the interaction and
pileup of magnetic flux tubes, and those regions provide the
possibility of generating strong field gradients, where particles can
experience substantial energization.
\\\\\\
C.A.G, P.D.M.,and  P.D. acknowledge support from grants UBACyT
No. 20020110200359 and 20020100100315, and from grants PICT
No. 2011-1529 and 2011-0454. W.H.M. was partially supported by NASA
LWS-TRT grant NNX15AB88G, Grand Challenge Research grant NNX14AI63G,
and the Solar Probe Plus mission through the Southwest Research
Institute ISIS project D99031L.
\bibliographystyle{plainnat}
\bibliography{Gonzalez_2017}

\providecommand{\noopsort}[1]{}\providecommand{\singleletter}[1]{#1}%
\begin{thebibliography}{31}
\providecommand{\natexlab}[1]{#1}
\providecommand{\url}[1]{\texttt{#1}}
\expandafter\ifx\csname urlstyle\endcsname\relax
  \providecommand{\doi}[1]{doi: #1}\else
  \providecommand{\doi}{doi: \begingroup \urlstyle{rm}\Url}\fi

\bibitem[Alexakis(2011)]{Alexakis2011}
Alexandros Alexakis.
\newblock Two-dimensional behavior of three-dimensional magnetohydrodynamic
  flow with a strong guiding field.
\newblock \emph{Phys. Rev. E}, 84:\penalty0 056330, Nov 2011.
\newblock \doi{10.1103/PhysRevE.84.056330}.
\newblock URL \url{http://link.aps.org/doi/10.1103/PhysRevE.84.056330}.

\bibitem[Ambrosiano et~al.(1988)Ambrosiano, Matthaeus, Goldstein, and
  Plante]{AmbrosianoEA88}
John Ambrosiano, William~H. Matthaeus, Melvyn~L. Goldstein, and Daniel Plante.
\newblock Test particle acceleration in turbulent reconnecting magnetic fields.
\newblock \emph{Journal of Geophysical Research: Space Physics}, 93\penalty0
  (A12):\penalty0 14383--14400, 1988.
\newblock ISSN 2156-2202.
\newblock \doi{10.1029/JA093iA12p14383}.
\newblock URL \url{http://dx.doi.org/10.1029/JA093iA12p14383}.

\bibitem[Chandran(2003)]{Chandran2003}
Benjamin D.~G. Chandran.
\newblock \emph{The Astrophysical Journal}, 599\penalty0 (2):\penalty0 1426,
  2003.
\newblock URL \url{http://stacks.iop.org/0004-637X/599/i=2/a=1426}.

\bibitem[Chandran and Maron(2004)]{CH1}
Benjamin D.~G. Chandran and Jason~L. Maron.
\newblock \emph{The Astrophysical Journal}, 603\penalty0 (1):\penalty0 23,
  2004.
\newblock URL \url{http://stacks.iop.org/0004-637X/603/i=1/a=23}.

\bibitem[Cho and Lazarian(2006)]{CHO1}
Jungyeon Cho and A.~Lazarian.
\newblock \emph{The Astrophysical Journal}, 638\penalty0 (2):\penalty0 811,
  2006.
\newblock URL \url{http://stacks.iop.org/0004-637X/638/i=2/a=811}.

\bibitem[Dalena et~al.(2012)Dalena, Greco, Rappazzo, Mace, and
  Matthaeus]{Dalena2012}
S.~Dalena, A.~Greco, A.~F. Rappazzo, R.~L. Mace, and W.~H. Matthaeus.
\newblock Magnetic moment nonconservation in magnetohydrodynamic turbulence
  models.
\newblock \emph{Phys. Rev. E}, 86:\penalty0 016402, Jul 2012.
\newblock \doi{10.1103/PhysRevE.86.016402}.
\newblock URL \url{https://link.aps.org/doi/10.1103/PhysRevE.86.016402}.

\bibitem[Dalena et~al.(2014)Dalena, Rappazzo, Dmitruk, Greco, and
  Matthaeus]{Dalena2014}
S.~Dalena, A.~F. Rappazzo, P.~Dmitruk, A.~Greco, and W.~H. Matthaeus.
\newblock \emph{The Astrophysical Journal}, 783\penalty0 (2):\penalty0 143,
  2014.
\newblock URL \url{http://stacks.iop.org/0004-637X/783/i=2/a=143}.

\bibitem[Dmitruk and Matthaeus(2006)]{hall}
P.~Dmitruk and W.~H. Matthaeus.
\newblock Test particle acceleration in three-dimensional hall mhd turbulence.
\newblock \emph{Journal of Geophysical Research: Space Physics}, 111\penalty0
  (A12):\penalty0 n/a--n/a, 2006.
\newblock ISSN 2156-2202.
\newblock \doi{10.1029/2006JA011988}.
\newblock URL \url{http://dx.doi.org/10.1029/2006JA011988}.
\newblock A12110.

\bibitem[Dmitruk et~al.(2004)Dmitruk, Matthaeus, and Seenu]{PD1}
Pablo Dmitruk, W.~H. Matthaeus, and N.~Seenu.
\newblock \emph{The Astrophysical Journal}, 617\penalty0 (1):\penalty0 667,
  2004.
\newblock URL \url{http://stacks.iop.org/0004-637X/617/i=1/a=667}.

\bibitem[Drake et~al.(2006)Drake, Swisdak, Che, and Shay]{DrakeEA06}
J.~F. Drake, M.~Swisdak, H.~Che, and M.~A. Shay.
\newblock Electron acceleration from contracting magnetic islands during
  reconnection.
\newblock \emph{Nature}, 443\penalty0 (7111):\penalty0 553--556, October 2006.
\newblock ISSN 0028-0836.
\newblock URL \url{http://dx.doi.org/10.1038/nature05116}.

\bibitem[Gonzalez et~al.(2016)Gonzalez, Dmitruk, Mininni, and
  Matthaeus]{Gonzalez2016_1}
C.~A. Gonzalez, P.~Dmitruk, P.~D. Mininni, and W.~H. Matthaeus.
\newblock On the compressibility effect in test particle acceleration by
  magnetohydrodynamic turbulence.
\newblock \emph{Physics of Plasmas}, 23\penalty0 (8):\penalty0 082305, 2016.
\newblock \doi{http://dx.doi.org/10.1063/1.4960681}.
\newblock URL
  \url{http://scitation.aip.org/content/aip/journal/pop/23/8/10.1063/1.4960681}.

\bibitem[Greco et~al.(2008)Greco, Chuychai, Matthaeus, Servidio, and
  Dmitruk]{Greco2008}
A.~Greco, P.~Chuychai, W.~H. Matthaeus, S.~Servidio, and P.~Dmitruk.
\newblock Intermittent mhd structures and classical discontinuities.
\newblock \emph{Geophysical Research Letters}, 35\penalty0 (19):\penalty0
  n/a--n/a, 2008.
\newblock ISSN 1944-8007.
\newblock \doi{10.1029/2008GL035454}.
\newblock URL \url{http://dx.doi.org/10.1029/2008GL035454}.
\newblock L19111.

\bibitem[Hussein and Shalchi(2016)]{hussein2016}
M.~Hussein and A.~Shalchi.
\newblock Simulations of energetic particles interacting with dynamical
  magnetic turbulence.
\newblock \emph{The Astrophysical Journal}, 817\penalty0 (2):\penalty0 136,
  2016.
\newblock URL \url{http://stacks.iop.org/0004-637X/817/i=2/a=136}.

\bibitem[{Karimabadi} et~al.(2013){Karimabadi}, {Roytershteyn}, {Wan},
  {Matthaeus}, {Daughton}, {Wu}, {Shay}, {Loring}, {Borovsky}, {Leonardis},
  {Chapman}, and {Nakamura}]{karimabadi2013}
H.~{Karimabadi}, V.~{Roytershteyn}, M.~{Wan}, W.~H. {Matthaeus}, W.~{Daughton},
  P.~{Wu}, M.~{Shay}, B.~{Loring}, J.~{Borovsky}, E.~{Leonardis}, S.~C.
  {Chapman}, and T.~K.~M. {Nakamura}.
\newblock {Coherent structures, intermittent turbulence, and dissipation in
  high-temperature plasmas}.
\newblock \emph{Physics of Plasmas}, 20\penalty0 (1):\penalty0 012303, January
  2013.
\newblock \doi{10.1063/1.4773205}.

\bibitem[Kulsrud.(1983)]{Kulsrud}
R.M. Kulsrud.
\newblock \emph{MHD description of plasma, in Basic Plasma Physics: Selected
  Chapter}.
\newblock AA Galeev, RN Sudan, 1983.

\bibitem[Lazarian et~al.(2012)Lazarian, Vlahos, Kowal, Yan, Beresnyak, and
  de~Gouveia Dal Pino]{L1}
A.~Lazarian, L.~Vlahos, G.~Kowal, H.~Yan, A.~Beresnyak, and E.M.
  de~Gouveia Dal Pino.
\newblock \emph{Space Science Reviews}, 173\penalty0 (1-4):\penalty0 557--622,
  2012.
\newblock ISSN 0038-6308.
\newblock \doi{10.1007/s11214-012-9936-7}.
\newblock URL \url{http://dx.doi.org/10.1007/s11214-012-9936-7}.

\bibitem[Lehe et~al.(2009)Lehe, Parrish, and Quataert]{lehe2009heating}
R{\'e}mi Lehe, Ian~J Parrish, and Eliot Quataert.
\newblock The heating of test particles in numerical simulations of
  alfv{\'e}nic turbulence.
\newblock \emph{The Astrophysical Journal}, 707\penalty0 (1):\penalty0 404,
  2009.

\bibitem[Lynn et~al.(2013)Lynn, Quataert, Chandran, and Parrish]{Lynn2013}
Jacob~W. Lynn, Eliot Quataert, Benjamin D.~G. Chandran, and Ian~J. Parrish.
\newblock \emph{The Astrophysical Journal}, 777\penalty0 (2):\penalty0 128,
  2013.
\newblock URL \url{http://stacks.iop.org/0004-637X/777/i=2/a=128}.

\bibitem[Matthaeus et~al.(1984)Matthaeus, Ambrosiano, and Goldstein]{M2}
W.~H. Matthaeus, J.~J. Ambrosiano, and M.~L. Goldstein.
\newblock \emph{Phys. Rev. Lett.}, 53:\penalty0 1449--1452, Oct 1984.
\newblock \doi{10.1103/PhysRevLett.53.1449}.
\newblock URL \url{http://link.aps.org/doi/10.1103/PhysRevLett.53.1449}.

\bibitem[McComas et~al.(2007)McComas, Velli, Lewis, Acton, Balat-Pichelin,
  Bothmer, Dirling, Feldman, Gloeckler, Habbal, Hassler, Mann, Matthaeus,
  McNutt, Mewaldt, Murphy, Ofman, Sittler, Smith, and Zurbuchen]{McCommas2007}
D.~J. McComas, M.~Velli, W.~S. Lewis, L.~W. Acton, M.~Balat-Pichelin,
  V.~Bothmer, R.~B. Dirling, W.~C. Feldman, G.~Gloeckler, S.~R. Habbal, D.~M.
  Hassler, I.~Mann, W.~H. Matthaeus, R.~L. McNutt, R.~A. Mewaldt, N.~Murphy,
  L.~Ofman, E.~C. Sittler, C.~W. Smith, and T.~H. Zurbuchen.
\newblock Understanding coronal heating and solar wind acceleration: Case for
  in situ near-sun measurements.
\newblock \emph{Reviews of Geophysics}, 45\penalty0 (1):\penalty0 n/a--n/a,
  2007.
\newblock ISSN 1944-9208.
\newblock \doi{10.1029/2006RG000195}.
\newblock URL \url{http://dx.doi.org/10.1029/2006RG000195}.
\newblock RG1004.

\bibitem[Mininni(2011)]{Mininni2011}
Pablo~D Mininni.
\newblock Scale interactions in magnetohydrodynamic turbulence.
\newblock \emph{Annual Review of Fluid Mechanics}, 43:\penalty0 377--397, 2011.

\bibitem[Mininni et~al.(2005)Mininni, Montgomery, and Pouquet]{Mininni2005}
Pablo~D Mininni, David~C Montgomery, and Annick~G Pouquet.
\newblock A numerical study of the alpha model for two-dimensional
  magnetohydrodynamic turbulent flows.
\newblock \emph{Physics of Fluids (1994-present)}, 17\penalty0 (3):\penalty0
  035112, 2005.

\bibitem[Mininni et~al.(2011)Mininni, Rosenberg, Reddy, and Pouquet]{ghost}
Pablo~D. Mininni, Duane Rosenberg, Raghu Reddy, and Annick Pouquet.
\newblock A hybrid mpi–openmp scheme for scalable parallel pseudospectral
  computations for fluid turbulence.
\newblock \emph{Parallel Computing}, 37\penalty0 (6–7):\penalty0 316 -- 326,
  2011.
\newblock ISSN 0167-8191.
\newblock \doi{http://dx.doi.org/10.1016/j.parco.2011.05.004}.
\newblock URL
  \url{http://www.sciencedirect.com/science/article/pii/S0167819111000512}.

\bibitem[Parker and Tidman(1958)]{Parker_1958}
E.~N. Parker and D.~A. Tidman.
\newblock Suprathermal particles.
\newblock \emph{Phys. Rev.}, 111:\penalty0 1206--1211, Sep 1958.
\newblock \doi{10.1103/PhysRev.111.1206}.
\newblock URL \url{http://link.aps.org/doi/10.1103/PhysRev.111.1206}.

\bibitem[Pouquet and Patterson(1978)]{pouquet1978}
A.~Pouquet and G.~S. Patterson.
\newblock Numerical simulation of helical magnetohydrodynamic turbulence.
\newblock \emph{Journal of Fluid Mechanics}, 85\penalty0 (2):\penalty0
  305--323, 03 1978.
\newblock \doi{10.1017/S0022112078000658}.

\bibitem[Servidio et~al.(2014)Servidio, Osman, Valentini, Perrone, Califano,
  Chapman, Matthaeus, and Veltri]{ServidioKinetic}
S.~Servidio, K.~T. Osman, F.~Valentini, D.~Perrone, F.~Califano, S.~Chapman,
  W.~H. Matthaeus, and P.~Veltri.
\newblock Proton kinetic effects in vlasov and solar wind turbulence.
\newblock \emph{The Astrophysical Journal Letters}, 781\penalty0 (2):\penalty0
  L27, 2014.
\newblock URL \url{http://stacks.iop.org/2041-8205/781/i=2/a=L27}.

\bibitem[Sujovolsky and Mininni(2016)]{sujovolski2016}
N.~E. Sujovolsky and P.~D. Mininni.
\newblock Tridimensional to bidimensional transition in magnetohydrodynamic
  turbulence with a guide field and kinetic helicity injection.
\newblock \emph{Phys. Rev. Fluids}, 1:\penalty0 054407, Sep 2016.
\newblock \doi{10.1103/PhysRevFluids.1.054407}.
\newblock URL \url{http://link.aps.org/doi/10.1103/PhysRevFluids.1.054407}.

\bibitem[Teaca et~al.(2014)Teaca, Weidl, Jenko, and Schlickeiser]{Bogdan2014}
Bogdan Teaca, Martin~S. Weidl, Frank Jenko, and Reinhard Schlickeiser.
\newblock \emph{Phys. Rev. E}, 90:\penalty0 021101, Aug 2014.
\newblock \doi{10.1103/PhysRevE.90.021101}.
\newblock URL \url{http://link.aps.org/doi/10.1103/PhysRevE.90.021101}.

\bibitem[Tessein et~al.(2015)Tessein, Ruffolo, Matthaeus, Wan, Giacalone, and
  Neugebauer]{Tessein2015}
Jeffrey~A. Tessein, David Ruffolo, William~H. Matthaeus, Minping Wan, Joe
  Giacalone, and Marcia Neugebauer.
\newblock Effect of coherent structures on energetic particle intensity in the
  solar wind at 1 au.
\newblock \emph{The Astrophysical Journal}, 812\penalty0 (1):\penalty0 68,
  2015.
\newblock URL \url{http://stacks.iop.org/0004-637X/812/i=1/a=68}.

\bibitem[Wareing and Hollerbach(2010)]{wareing2010}
Christopher~J. Wareing and Rainer Hollerbach.
\newblock Cascades in decaying three-dimensional electron magnetohydrodynamic
  turbulence.
\newblock \emph{Journal of Plasma Physics}, 76\penalty0 (1):\penalty0 117--128,
  002 2010.
\newblock \doi{10.1017/S0022377809990158}.

\bibitem[Weidl et~al.(2015)Weidl, Jenko, Teaca, and Schlickeiser]{Weidl2015}
Martin~S. Weidl, Frank Jenko, Bogdan Teaca, and Reinhard Schlickeiser.
\newblock \emph{The Astrophysical Journal}, 811\penalty0 (1):\penalty0 8, 2015.
\newblock URL \url{http://stacks.iop.org/0004-637X/811/i=1/a=8}.

\end{thebibliography}


\end{document}